\documentclass[sigconf, nonacm, pdfa]{acmart}

\usepackage[ruled, linesnumbered]{algorithm2e}
\SetKwInOut{Input}{Input}
\SetKwInOut{Output}{Output}
\SetKwComment{Comment}{/* }{ */}        
\usepackage{enumitem}                   
\usepackage[framemethod=tikz]{mdframed} 
\usepackage{graphicx}                   
\usepackage{colortbl}
\usepackage{hyperref}                   
\hypersetup{
  colorlinks=true,
  linkcolor=blue,
  urlcolor=cyan,
}
\usepackage{multirow}                   
\usepackage[a-2b]{pdfx}                 
\usepackage{xcolor}                     
\usepackage{xspace}                     
\usepackage{mathtools}
\usepackage{balance}

\newcommand\vldbdoi{10.14778/3636218.3636237}
\newcommand\vldbpages{849 - 862}
\newcommand\vldbvolume{17}
\newcommand\vldbissue{4}
\newcommand\vldbyear{2023}
\newcommand\vldbauthors{\authors}
\newcommand\vldbtitle{\shorttitle} 
\newcommand\vldbavailabilityurl{URL_TO_YOUR_ARTIFACTS}
\newcommand\vldbpagestyle{empty}

\newcommand{\sysname}{\mbox{\textsc{Observatory}}\xspace}

\newcommand{\bert}{\texttt{BERT}\xspace}
\newcommand{\roberta}{\texttt{RoBERTa}\xspace}
\newcommand{\tfive}{\texttt{T5}\xspace}
\newcommand{\turl}{\texttt{TURL}\xspace}
\newcommand{\doduo}{\texttt{DODUO}\xspace}
\newcommand{\tapas}{\texttt{TAPAS}\xspace}
\newcommand{\tabert}{\texttt{TaBERT}\xspace}
\newcommand{\tapex}{\texttt{TaPEx}\xspace}
\newcommand{\taptap}{\texttt{TapTap}\xspace}
\renewcommand{\paragraph}[1]{\vspace{0.2\baselineskip}\noindent\textbf{#1.}}

\theoremstyle{definition}
\newtheorem{definition}{Definition}
\newtheorem{property}{Property}
\newtheorem{measure}{Measure}

\newtheorem{mdexample}{\hspace{5pt}Example}

\newtheorem*{mdexample*}{Example}
\newenvironment{example*}{\begin{mdframed}[backgroundcolor=teal!12, roundcorner=10pt, linewidth=0pt, innertopmargin=1pt, innerbottommargin=5pt, skipabove=5pt, skipbelow=3pt]\begin{mdexample*}}{\end{mdexample*}\end{mdframed}}

\begin{document}
\title{\sysname: Characterizing Embeddings of Relational Tables}

\author{Tianji Cong}
\affiliation{%
  \institution{University of Michigan}
}
\email{congtj@umich.edu}

\author{Madelon Hulsebos}
\affiliation{%
  \institution{University of Amsterdam}
}
\email{m.hulsebos@uva.nl}

\author{Zhenjie Sun}
\affiliation{
  \institution{University of Michigan}
}
\email{zjsun@umich.edu}

\author{Paul Groth}
\affiliation{%
  \institution{University of Amsterdam}
}
\email{p.t.groth@uva.nl}

\author{H. V. Jagadish}
\affiliation{
  \institution{University of Michigan}
}
\email{jag@umich.edu}

\begin{abstract}
  Language models and specialized table embedding models have recently demonstrated strong performance on many tasks over tabular data. Researchers and practitioners are keen to leverage these models in many new application contexts; but limited understanding of the strengths and weaknesses of these models, and the table representations they generate, makes the process of finding a suitable model for a given task reliant on trial and error. There is an urgent need to gain a comprehensive understanding of these models to minimize inefficiency and failures in downstream usage. 
  
  To address this need, we propose \sysname, a formal framework to systematically analyze embedding representations of relational tables. Motivated both by invariants of the relational data model and by statistical considerations regarding data distributions, we define eight primitive properties, and corresponding measures to quantitatively characterize table embeddings for these properties. 
  Based on these properties, we define an extensible framework to evaluate language and table embedding models. We collect and synthesize a suite of datasets and use \sysname to analyze nine such models. Our analysis provides insights into the strengths and weaknesses of learned representations over tables. We find, for example, that some models are sensitive to table structure such as column order, that functional dependencies are rarely reflected in embeddings, and that specialized table embedding models have relatively lower sample fidelity. Such insights help researchers and practitioners better anticipate model behaviors and select appropriate models for their downstream tasks, while guiding researchers in the development of new models.
\end{abstract}

\maketitle

\pagestyle{\vldbpagestyle}
\begingroup\small\noindent\raggedright\textbf{PVLDB Reference Format:}\\
\vldbauthors. \vldbtitle. PVLDB, \vldbvolume(\vldbissue): \vldbpages, \vldbyear.\\
\href{https://doi.org/\vldbdoi}{doi:\vldbdoi}
\endgroup
\begingroup
\renewcommand\thefootnote{}\footnote{\noindent
This work is licensed under the Creative Commons BY-NC-ND 4.0 International License. Visit \url{https://creativecommons.org/licenses/by-nc-nd/4.0/} to view a copy of this license. For any use beyond those covered by this license, obtain permission by emailing \href{mailto:info@vldb.org}{info@vldb.org}. Copyright is held by the owner/author(s). Publication rights licensed to the VLDB Endowment. \\
\raggedright Proceedings of the VLDB Endowment, Vol. \vldbvolume, No. \vldbissue\ %
ISSN 2150-8097. \\
\href{https://doi.org/\vldbdoi}{doi:\vldbdoi} \\
}\addtocounter{footnote}{-1}\endgroup

\ifdefempty{\vldbavailabilityurl}{}{
\begingroup\small\noindent\raggedright\textbf{PVLDB Artifact Availability:}\\
The code and data are available at \url{https://github.com/superctj/observatory}.
\endgroup
}

\section{Introduction}
  \begin{figure*}
    \centering
    \includegraphics[width=0.9\textwidth]{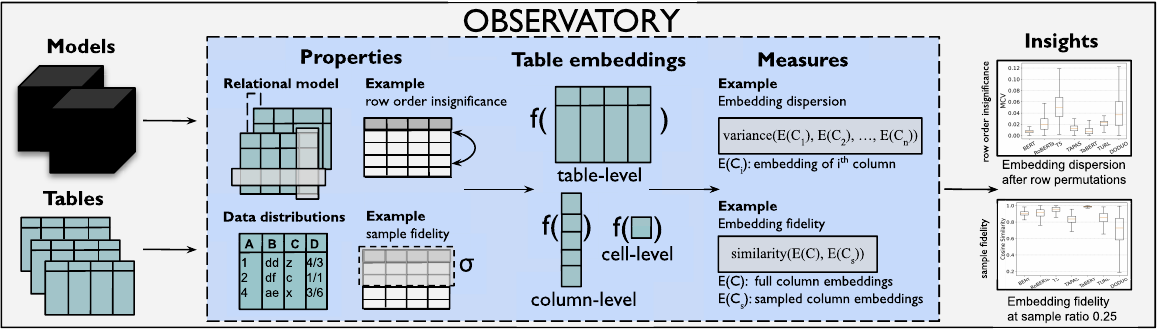}
    \caption{Overview of \sysname and how it solicits understanding of opaque table embedding models by measuring properties motivated by the relational data model and data distributions. We illustrate the framework for two out of eight properties: 1) row order insignificance, and 2) sample fidelity.}
    \label{fig:overview}
  \end{figure*}

  The advances of pretrained language models for NLP tasks such as summarization and dialog have sparked similar interest and progress in embedding relational tables for tasks such as table question answering~\cite{DBLP:conf/acl/HerzigNMPE20}, entity matching~\cite{DBLP:journals/pvldb/DengSL0020, DBLP:journals/pvldb/0001LSDT20}, semantic column type annotation~\cite{DBLP:journals/pvldb/ZhangSLHDT20, DBLP:journals/pvldb/DengSL0020, DBLP:conf/sigmod/SuharaL0ZDCT22}, and data integration and augmentation~\cite{DBLP:conf/sigmod/CappuzzoPT20, DBLP:journals/pvldb/TangFLTDLMO21, DBLP:journals/corr/cong2023pylon}. Most of these models are built on top of language models such as BERT~\cite{DBLP:conf/naacl/DevlinCLT19} and specialized to take into account the structure of tables, for example, by leveraging vertical attention to incorporate information across rows~\cite{DBLP:conf/acl/YinNYR20}.

  As these table embedding models have shown strong performance on a variety of tasks, researchers and practitioners are also interested in using these pretrained models for new applications and in new domains. However, the process of identifying a suitable model typically involves trial and error due to a lack of understanding regarding the strengths and limitations of these models and their learned representations. This knowledge gap can produce inefficiency and even failures in downstream usage. Moreover, researchers have little visibility into the behaviors and generalizability of existing table embedding models beyond their performance on particular downstream tasks. Hence, there is a pressing need to understand the strengths and weaknesses of these models, especially in terms of the table embeddings they generate~\cite{DBLP:conf/ijcai/dong22, badaro2023transformers}.
  
  To address this need, we propose \sysname, a formal framework for systematically analyzing language- and table embedding models from the perspective of what characteristics of relational tables these models do and do not capture in their learned embedding representations. \sysname presents eight primitive properties motivated both by invariants in Codd's relational data model~\cite{DBLP:journals/cacm/Codd70, DBLP:journals/tods/Codd79} and by statistical considerations regarding data distributions in downstream tasks: for instance, if embeddings are sensitive to row and column order or sample size. Each of these properties is associated with a measure that quantitatively characterizes embedding representations over relational tables (see Figure~\ref{fig:overview} for an overview). Analogous to task-agnostic analyses of language models~\cite{DBLP:conf/acl/RibeiroWGS20, DBLP:journals/corr/AdiKBLG16}, such data-specific evaluations of embeddings offer valuable insights into model behaviors, which are connected to various downstream applications (Section~\ref{sec:conn} gives more details). With \sysname, we 1) consolidate properties of relational tables important to reflect in table embeddings, 2) contribute a framework and implementation thereof, enabling researchers and practitioners to analyze the capabilities of existing and new models with respect to these properties, and 3) provide insights into the strengths and limitations of nine popular models through their learned representations over tabular data, which can inform researchers and practitioners of model selections and novel model designs. 

  Along with the implementation of \sysname, we collect and synthesize a suite of datasets for evaluation purposes, and present a comprehensive analysis of nine commonly used language and specialized table embedding models. Some key insights we surface in our analysis are that the embeddings of some models are sensitive to the order of rows and, in particular, the order of columns, while embeddings of some models are robust to uniform sampling. Moreover, we find that none of the models reflect functional dependencies among columns in tables. Although we do not aim to, and cannot, analyze all existing models, our implementation of \sysname is extensible such that researchers and practitioners can use \sysname for analysis of new models by specifying the procedure of embedding inference following the implemented interface. In summary, we make the following contributions:
  \begin{itemize}[leftmargin=10pt]
    \item We propose \sysname, a framework including eight primitive properties and corresponding measures for systematically analyzing embedding representations over relational tables.
    \item We implement and open-source a prototype of \sysname, which covers nine popular table embedding models while also being extensible for evaluation of new models.
    \item We present a comprehensive analysis with \sysname and provide novel insights into the strengths and limitations of evaluated models and their learned table representations.
  \end{itemize}

\section{Related Work}
\subsection{Language and Table Embedding Models}


\paragraph{Language Models} BERT~\cite{DBLP:conf/naacl/DevlinCLT19} is among the first transformer-based pretrained language models, generating contextual embeddings by predicting masked tokens. Subsequent optimizations like RoBERTa~\cite{DBLP:journals/corr/liu19} and expansions in model size and tasks (e.g., T5~\cite{DBLP:journals/jmlr/RaffelSRLNMZLL20}) have driven rapid advancements. Language models soon progress from predictive tasks to sequence generation, exemplified by GPT models~\cite{radford2018improving}. Beyond unstructured language tasks, investigations explore language models' capabilities for structured inputs like tabular data. Narayan et al.~\cite{DBLP:journals/pvldb/NarayanCOR22} uses T5 for data wrangling, and recent GPT-based conversational models directly handle table understanding tasks\cite{DBLP:journals/corr/korini2023column, DBLP:journals/corr/kayali2023chorus}.


\paragraph{Table Embedding Models} TaBERT~\cite{DBLP:conf/acl/YinNYR20} pioneers extending pretrained language models to tabular data. It employs token-level embeddings with additional positional embeddings, incorporating vertical attention for inter-row information and a masked column name prediction objective inspired by BERT. Subsequent models, including TURL~\cite{DBLP:journals/pvldb/DengSL0020}, TAPAS~\cite{DBLP:conf/acl/HerzigNMPE20}, and TaPEx~\cite{DBLP:conf/iclr/LiuCGZLCL22}, facilitate applications like table question answering, table understanding, and data preparation. For a comprehensive overview, we refer readers to surveys by Dong et al.~\cite{DBLP:conf/ijcai/dong22} and  Badaro et al.~\cite{badaro2023transformers}. The latter emphasizes the need for intrinsic analysis of table embedding models, which we take a first step towards addressing with \sysname.

\subsection{Analysis of Embedding Models} 

\paragraph{Analysis of Language Embedding Models} 
Efforts to comprehend and evaluate LMs involve task-specific~\cite{DBLP:conf/iclr/WangSMHLB19} and task-agnostic analyses~\cite{DBLP:conf/acl/RibeiroWGS20}. Task-agnostic investigations, exemplified by CheckList~\cite{DBLP:conf/acl/RibeiroWGS20}, explore internal LM behavior and capacities through unit-test-like assessments (e.g., whether a LM can handle negation). In line with this, \sysname proposes relational data model-inspired properties, considering practical data distribution factors for downstream applications. Recently, Sui et al.~\cite{DBLP:journals/corr/sui2023structural} introduces a benchmark evaluating LMs on seven table tasks (e.g., cell lookup) while varying, among others, prompt designs and table input formatting. However, it falls short in examining fundamental properties of relational tables and data distributions, and excluding specialized table embedding models.


\paragraph{Analysis of Table Embedding Models} Limited analyses exist on table embedding models. Wang et al.~\cite{DBLP:journals/corr/WangJN022} assess the impact of explicitly modeling table structure in transformer architectures for table retrieval, revealing the modest contribution of table-specific model design. However, this evaluation is confined to retrieval tasks and lacks insights into intrinsic model limitations affecting downstream performance. Dr.Spider~\cite{DBLP:journals/corr/chang2023drspider} benchmarks text-to-SQL models for perturbation robustness, while \sysname introduces novel properties, including perturbation robustness, unexplored until now. Recent work~\cite{DBLP:journals/corr/SrinivasDAHK0PCS23} introduces LakeBench, highlighting performance gaps in specialized table embedding models for data discovery. In contrast, \sysname evaluates embedding representations based on broader table-specific properties relevant to diverse downstream tasks. Koleva et al.~\cite{koleva2022analysis} examines patterns in table-specific attention mechanisms, remaining task-agnostic. Unlike \sysname, it doesn't link model analysis with relational and data distribution properties of tables.

\section{\sysname}
  In this section, we present \sysname, our methodology for characterizing embedding representations over relational tables. \sysname features two sets of properties that are agnostic to downstream tasks and motivated by the relational model~\cite{DBLP:journals/cacm/Codd70, DBLP:journals/tods/Codd79} and data distributions. For each property, \sysname proposes a measure to quantify how well embedding representations align with the property specification. This allows users to gain insights into the strengths and weaknesses of different models and to even compare models through a consistent lens.

\subsection{Problem Statement}\label{sec:methodol_ps}
  Various downstream applications may need different kinds of embeddings. For example, semantic column type detection is based on column embeddings whereas entity matching requires entity embeddings. Given that these embeddings look at different {\em levels of aggregation} of the table structure, we refer to these kinds of embeddings as levels of embeddings.
  \begin{definition}[Table Embedding Characterization]
    Given a pretrained model $f$, a corpus of tables $T \in \mathcal{T}$, and a property $\mathcal{P}$ that characterizes a certain level of embeddings $\mathbf{E_{\mathcal{P}}}$ with a measure $\mathcal{M}$, table embedding characterization infers $\mathbf{E_{\mathcal{P}}}$ with $f$ over each $T \in \mathcal{T}$ and computes $\mathcal{M}$ over the distribution of $\mathbf{E_{\mathcal{P}}}$.
  \end{definition}
  A property $\mathcal{P}$ can characterize one or more levels of embeddings (e.g., it can apply for both row- and column-level embeddings). Properties in \sysname span five levels of embeddings: table, column, row, cell, and entity (so called \textit{table embeddings}) while many of them are relevant to column-level embeddings. \sysname also focuses on Transformer-based embedding models. Technically, any pretrained model $f$, regardless of the architecture (encoder or encoder-decoder or decoder-only) can be integrated to and evaluated with \sysname, as long as $f$ either natively exposes certain level of embeddings $\mathbf{E_{\mathcal{P}}}$ specified by $\mathcal{P}$ or exposes token-level embeddings that can be further aggregated to the level of $\mathbf{E_{\mathcal{P}}}$.

\subsection{Relational Properties}
  The relational data model specifies both structural invariants and semantics. We first introduce two properties from structural invariants (namely, Row- and Column Order Insignificance), followed by two properties from structural semantics (namely, Join Relationship and Functional Dependencies).

  \begin{property}[Row Order Insignificance]\label{prop:row_order}
    A relational table can be viewed as a set of rows of which, in principle, the order is insignificant~\cite{DBLP:journals/cacm/Codd70}. Tables may be stored in an ordered way, that is, rows may be ordered by dates, or ascending/descending values of a given column. Models that explicitly encode the table structure with position embeddings might reflect this order in the output embeddings. Awareness of the influence of row order on table embeddings is key to using them in a context of unordered tables. We consider column/row/table-level embeddings in this property.
    
    \begin{measure}\label{measure:row_order}
      Given a table $T$, let $\mathbf{E}(D^{(i)})$ denote the embedding of column/row/table $D$ in the $i$-th row-wise shuffle of $T$ for $\leq i \leq n$ (i.e., there are $n$ row-wise permutations). We define the row order sensitivity as a high-dimensional dispersion measure $\mathcal{M}$ of $n$ samples drawn from the embedding distribution, i.e., $$\mathcal{M}(\; \mathbf{E}(D^{(1)}), \mathbf{E}(D^{(2)}), \dots, \mathbf{E}(D^{(n)}) \;).$$
    
      The coefficient of variation (CV), the ratio of the standard deviation to the mean in the univariate setting, is a well-known measure of variability relative to the mean of a population. It has the merit of allowing for the comparison of random variables with different units or different means. Thus, we consider multivariate extensions of CV (MCV) that summarize relative variation of a random vector (instead of a random variable) into a scalar quantity. In particular, we use Albert and Zhang's MCV~\cite{albert2010novel} to compare row order sensitivity across models for the reasons that it takes into account correlations between variables and does not require the covariance matrix to have an inverse~\cite{albert2010novel, DBLP:journals/ma/AertsHR15}, which is especially convenient when the number of observations (number of embeddings) is smaller than the number of variables (embedding dimensionality). Albert and Zhang's MCV of embeddings $\{ \mathbf{E}(C^{(i)})\}_{i=1}^{n}$ is computed as
      \begin{equation}\label{eqn:mcv}
        \gamma_{AZ} = \sqrt{ \frac{\mu^{t} \Sigma \mu}{(\mu^{t} \mu)^{2}} }
      \end{equation}
      where $\mu$ is the mean vector and $\Sigma$ is the covariance matrix.
    \end{measure}

    In practice, the number of possible permutations can be large (i.e., factorial of the number of rows) for tables with high cardinality. For computational efficiency in the experiments, we use at most 1000 randomly generated permutations of each table.

    \begin{example*}
      Figure~\ref{fig:row_permutations} gives an example of row permutations. Given 6 data rows, there are in total $6!= 720$ possible permutations. Then for each column, we have $720$ observations of embeddings, which is smaller than some embedding dimensionality (e.g., 768 of \bert). In this case, the covariance matrix derived from observations is singular. Nevertheless, Albert and Zhang's MCV can be calculated whereas other MCVs surveyed in~\cite{DBLP:journals/ma/AertsHR15} can not.
    \end{example*}

    \begin{figure}[ht!]
      \centering
      \includegraphics[width=0.95\linewidth]{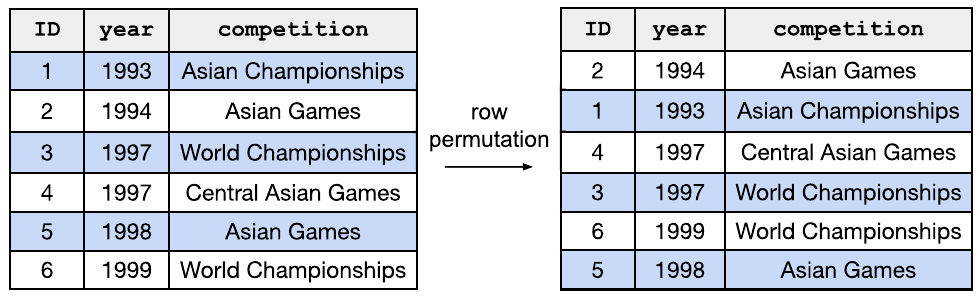}
      \caption{Illustration of row permutations.} 
      \label{fig:row_permutations}
    \end{figure}
  \end{property}

  \begin{property}[Column Order Insignificance]\label{prop:col_order}
    Besides row order, some models exploit neighboring columns as context when learning representations based on the intuition that neighboring columns can provide local context~\cite{DBLP:journals/pvldb/ZhangSLHDT20, DBLP:journals/corr/abs-2305-09696}. Analogous to row order insignificance, relational tables usually store data without preserving a particular column order.
    The (in)sensitivity of embeddings regarding the column order informs their suitability for tasks such as join discovery and table understanding in relational databases with unordered tables versus views on Web and other media that may present data with related attributes next to each other. As in Property~\ref{prop:row_order}, we assess column/row/table embeddings.

    \begin{measure}
      Given a table $T$, let $\mathbf{E}(D^{(i)})$ be the embedding of column/row/table $D$ in the $i$-th column-wise shuffle of $T$. Similarly, we measure the embedding variance using MCV in equation~\ref{eqn:mcv}.
    \end{measure}
  \end{property}

  \begin{property}[Join Relationship]\label{prop:join}
    The join operation combining tuples from two or more relational tables, is one of the essential operations for data analysis. Thus the problem of finding join candidates in a table repository has been extensively studied~\cite{DBLP:journals/pvldb/ZhangHOPS10, DBLP:conf/sigmod/YakoutGCC12, DBLP:journals/pvldb/ZhuNPM16, DBLP:conf/sigmod/ZhuDNM19, DBLP:conf/edbt/0002N021, DBLP:conf/cidr/CongGFJD23}. Join candidates are typically identified by some notion of value overlap similarity such as Jaccard and containment~\cite{DBLP:journals/pvldb/ZhuNPM16, DBLP:conf/sigmod/ZhuDNM19, DBLP:conf/edbt/0002N021} while the embedding approach has also been explored~\cite{DBLP:conf/cidr/CongGFJD23}. Their findings indicate that columns with significant value overlap are also close to each other in the embedding space. We investigate this postulate by assessing if there is a monotonic relationship between value overlap and embedding similarity. 

    \begin{measure}
      Consider pairs of query and candidate columns $(C_{q}, \\C_{c})$ and their corresponding embeddings $(\mathbf{E}({C_{q}}), \mathbf{E}({C_{c}}))$. Two random variables can be derived, the embedding similarity measure $\mathcal{M}(\mathbf{E}({C_{q}}), \mathbf{E}({C_{c}}))$ and the value overlap measure $\mathcal{R}(C_{q}, C_{c})$. In experiments, we use cosine similarity for $\mathcal{M}$ and containment for $\mathcal{R}$ where $\mathcal{R} = \frac{\vert C_{q} \cap C_{c} \vert}{\vert C_{q} \vert}$ and is not biased towards small sets~\cite{DBLP:journals/pvldb/ZhuNPM16, DBLP:conf/sigmod/ZhuDNM19}. For completeness, we also experiment with Jaccard similarity (i.e., $\frac{\vert C_{q} \cap C_{c} \vert}{\vert C_{q} \cup C_{c} \vert}$) and multiset Jaccard similarity (i.e., $\frac{\vert C_{q} \cap C_{c} \vert}{\vert C_{q} \vert + \vert C_{c} \vert}$) for measuring value overlap.
      
      With embedding similarity measure $\mathcal{M}$ and value overlap measure $\mathcal{R}$ calculated over $n$ pairs of query and candidate columns $\{(M_{1}, R_{1}), (M_{2}, R_{2}), \dots, (M_{n}, R_{n})\}$, we compute the Spearman's \\rank correlation coefficient between $\mathcal{M}$ and $\mathcal{R}$ as 
      \begin{equation}\label{eqn:spearman}
        \rho = \frac{\text{cov}(R(\mathcal{M}), R(\mathcal{R}))}{\sigma_{R(\mathcal{M})} \sigma_{R(\mathcal{R})}}
      \end{equation} where $R(\cdot)$ denotes the rank of a sample, $\text{cov}(\cdot, \cdot)$ is the covariance of the rank variables, and $\sigma_{(\cdot)}$ denotes the standard deviation.
  
      Note that the Spearman coefficient ranges between -1 and 1, and considers the ranking values of two variables instead of raw variable values. A coefficient of 1 means the rankings of each variable match up for all data pairs and indicates there is a very strong positive monotonic relationship between two variables. We adopt the Spearman coefficient since it does not make any assumption of the underlying variable distributions.  
    \end{measure}
  \end{property}

  \begin{property}[Functional Dependencies]\label{prop:fd}
    Let $T$ be a relation with a set of attributes $U$. Relation $T$ over $U$ is said to satisfy a functional dependency, denoted $T \models X \rightarrow Y$ where $X, Y \subset U$, if for each pair $s, t$ of tuples in $T$, $\pi_{X}(s) = \pi_{X}(t)$ implies $\pi_{Y}(s) = \pi_{Y}(t)$~\cite{DBLP:books/aw/AbiteboulHV95}. Functional dependencies between columns provide a formal mechanism to express semantic constraints to the stored data, which is useful in many applications such as improving schema design, data imputation, and query optimization.
    
    This property surfaces if models implicitly capture the relationship of functional dependencies in their representations (we are not aware of any model that explicitly takes functional dependencies into consideration in pretraining). Analogous to relationships between words~\cite{DBLP:conf/nips/MikolovSCCD13} and entities in knowledge bases~\cite{DBLP:conf/nips/BordesUGWY13}, the functional dependency relationship can be interpreted as a translation in the embedding space. Consider the relation triple ($\pi_{X}(s)$, \textit{r}, $\pi_{Y}(s)$), where \textit{r} is the functional dependent relationship between value pair $\pi_{X}(s)$, $\pi_{Y}(s)$. As demonstrated in~\cite{DBLP:conf/nips/BordesUGWY13}, such relationship reflects as a \textit{translation} between 
    the embeddings $\mathbf{E}(\pi_{X}(s))$ and $\mathbf{E}(\pi_{Y}(s))$. The translation vector represents relationship $r$, which can be expected to remain equal in direction and magnitude across tuples if the relationship is preserved~\cite{DBLP:conf/nips/BordesUGWY13, DBLP:conf/nips/MikolovSCCD13}. More precisely, consider any pair $s, t$ of tuples in $T$ with a functional dependency $X \rightarrow Y$. We say that this functional dependency is preserved in an embedding space determined by a model $f$ if $$d(\; \mathbf{E}(\pi_{X}(s)),\; \mathbf{E}(\pi_{Y}(s)) \;) = d(\; \mathbf{E}(\pi_{X}(t)),\; \mathbf{E}(\pi_{Y}(t)) \;)$$ given $\pi_{X}(s) = \pi_{X}(t)$ where $\mathbf{E}(\cdot)$ is the embedding inferred with $f$ and $d$ denotes a distance metric preserving direction and magnitude.

    \begin{example*}
      Consider a table $T$ containing four columns in Figure~\ref{fig:fd_table}. There exists a functional dependency between non-key attributes \texttt{country} and \texttt{continent}, i.e., \texttt{country} $\rightarrow$ \\\texttt{continent}. $T$ satisfies this functional dependency because every instance of a specific value in column \texttt{country}, \textit{Netherlands} for example, corresponds to the same value, i.e. \textit{Europe}, in the corresponding tuples under column \texttt{continent}. By our definition, if an embedding space preserves functional dependencies, the squared Euclidean distances between embeddings generated for these specific value pairs will be (approximately) equal, despite influence of context on the embeddings.
    \end{example*}
    
    \begin{figure}[!ht]
      \includegraphics[width=0.7\linewidth]{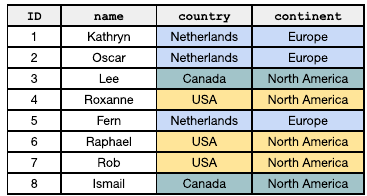}
      \caption{Table with a functional dependency \texttt{country} $\rightarrow$ \texttt{continent}. The colors illustrate different FD groups determined by the unique values in the \texttt{country} column.}
      \label{fig:fd_table}
    \end{figure}

    \begin{measure}
      Given a table $T$ with functional dependency $X \rightarrow Y$, we refer to the group of tuples $\pi_{X \cup Y}$ with the same value $v_{X}$ of determinant $X$ as FD-group $\mathcal{G}_{v_X}$, to the value associated with $v_X$ in the dependent attribute set $Y$ as $v_Y$, and to the embeddings of these values of the $i$-th entry in the group as $\mathbf{E}(v_{X, i})$ and $\mathbf{E}(v_{Y, i})$, respectively. For instance, there are three FD-groups under the functional dependency \texttt{country} $\rightarrow$ \texttt{continent} in the table shown in Figure~\ref{fig:fd_table}, i.e., (Netherlands, Europe), (Canada, North America), (USA, North America) where the FD-group (Netherlands, Europe) has three entries.

      Within each FD-group $\mathcal{G}_{j}$ of size $m_{\mathcal{G}_{j}}$, we calculate distance metric $d$ for each embedding pair ($\mathbf{E}(v_{X, i})$, $\mathbf{E}(v_{Y, i})$), denoted as $d_{ji}$.The average group-wise variance over all $n$ FD-groups is calculated as: $$\overline{S^2} = \frac{1}{n} \sum_{j=1}^{n} \frac{\sum^{m_{\mathcal{G}_j}}_{i=1}{||d_{ji} - \overline{d_{j}}||_{2}^2}}{m_{\mathcal{G}_{j}}-1}$$
    
      In our experiments, we take as distance metric $d$ the $L_{1}$- or $L_{2}$-norm following~\cite{DBLP:conf/nips/BordesUGWY13}, while other distance metrics preserving norm direction and magnitude are valid too. $\overline{S^2}$ approaches 0 if the \textit{translation} between the group-wise FD value pairs in $X$ (\texttt{country}) and $Y$ (\texttt{continent}) remains approximately equal for each FD group. We note that this does not require a strictly injective model. That is, the same value across different table contexts is not necessarily mapped to exactly the same vector in the embedding space in order for this measure to approach 0.

      In addition, it is expected that this measure shows higher value ranges over column sets without functional dependencies. We collect a set of functional dependencies over tables $\mathcal{T}_{FD}$ and a set of tables $\mathcal{T}_{\neg FD}$ in which no table contains functional dependent columns. We calculate the measure for all tables in the sets $\mathcal{T}_{FD}$ and $\mathcal{T}_{\neg FD}$. This yields two distributions of $\overline{S^2}$ values. If the embeddings preserve functional dependencies, $\overline{S^2}$ values over $\mathcal{T}_{FD}$ will be close to 0 and in general smaller than those over $\mathcal{T}_{\neg FD}$.
    \end{measure}
  \end{property}

\subsection{Data Distribution Properties}
  In practice, many aspects need to be considered when using embeddings including but not limited to the sample size, domain generalizability, robust representations of semantically similar values, and context. We introduce four properties involving data distributions that concern these four aspects.

  \begin{property}[Sample Fidelity]\label{prop:sample_fidelity}
    Large relational tables can easily have millions or even billions of rows. Embedding an entire table or even a single large column with a model is often infeasible due to constraints on the input length of models or memory constraints of computing resources. On the other hand, it may not be necessary to embed the full table for a downstream task~\cite{DBLP:conf/acl/YinNYR20, DBLP:conf/sigmod/SuharaL0ZDCT22, DBLP:conf/cidr/CongGFJD23}. In practice, existing work resorts to sampling, either up to the input limit or based on content relevance, as a straightforward workaround. While sampling provides a feasible solution, it also introduces a trade-off between computational cost and the fidelity of the embedding inferred from a smaller sample compared to the embedding that would have been obtained if the entire dataset were used. It is then essential to understand the fidelity of sample embeddings from a model by evaluating the extent to which sample embeddings deviate from the embeddings of full values.

    \begin{measure}
      Given a full column $C$ and a sample $C_{\mathcal{S}}$, we define sample fidelity as a similarity measure $\mathcal{M}$ between the embedding of the full column $\mathbf{E}(C)$ and the sample embedding $\mathbf{E}(C_{\mathcal{S}})$ where $\mathcal{M}$ can be cosine similarity for instance. Similar to~\cite{DBLP:conf/kdd/WangDJLFHZ21}, we split a full column into chunks with the shared header and obtain the full embedding by aggregating the chunk embeddings. This is because a full column may not fit into a single sequence for model ingestion.
      
      For each column $C$, we perform uniform random sampling to get $n$ distinct samples $\{ C_{1}, C_{2}, \dots, C_{n} \}$ from $C$ and report the average column sample fidelity $$\frac{1}{n} \sum_{i=1}^{n} \mathcal{M}(\mathbf{E}(C_{i}), \mathbf{E}(C_{i \mathcal{S}}))$$ as well as the multivariate coefficient of variation over the embedding set $\{\mathbf{E}(C), \mathbf{E}(C_{1}), ..., \mathbf{E}(C_{n}) \}$. Since tables in a corpus may have various sizes, we experiment with different sampling fractions (e.g., 0.25, 0.5, and 0.75) instead of varying the absolute number of samples in evaluations.
    \end{measure}

    This simple measure gives a good indication of computing efficiency and monetary cost. For example, provided that cloud vendors take a pay-as-you-go model, users do not need to pull out all their data to infer embeddings and pay the full scanning cost.
  \end{property}

  \begin{property}[Entity Stability]\label{prop:entity_stb}
    Stability is a notion in NLP~\cite{DBLP:conf/naacl/WendlandtKM18, DBLP:journals/tacl/AntoniakM18} that indicates the variability of word embeddings relative to training data, training algorithms, and other factors in embedding model training. The idea is to use the overlap between $K$ nearest neighbors of queries (i.e., words) found in different embedding spaces\footnote{An embedding space refers to a vector space that represents an original space of inputs (e.g., words or table columns).} as a proxy of agreement between embedding spaces. We borrow this notion to explore the (in)stability of entity embeddings. 

    Given $n$ embedding spaces determined by embedding models ${\mathbf{f}_{1}, \mathbf{f}_{2}, \dots, \mathbf{f}_{n}}$, consider an entity cell $\mathbf{e}=(e_{m}, e_{md})$ in a relational table where $e_{m}$ is the entity mention and $e_{md}$ is associated metadata if exist (such as the entity linked to the cell from a knowledge base, the column name, and the table caption). Retrieve $K$ nearest neighbor entities of $\mathbf{e}$ in each embedding space. The stability of entity $\mathbf{e}$ across $n$ embedding spaces is defined as the average over all pairwise percent overlap between two embedding spaces. 

    \begin{example*}
      Take the entity column \textit{competition} in Figure~\ref{fig:row_permutations} for example. \textit{World\ Championships} is an entity mention that links to a Wikipedia entity \textit{1997\_World\_Championships\_in\_Athletics\\\_-\_Men's\_Decathlon}. Depending on the context, the same entity mention may link to another distinct entity, for instance, \textit{BWF\_World\_Championships}.
    \end{example*}

    \begin{measure}
      We consider the case when $n=2$ (i.e., two embedding models $\mathbf{f}_{1}$ and $\mathbf{f}_{2}$). We randomly sample $m$ entities, and for each entity $\mathbf{e_{i}}$, let $\mathit{s^{i}_{1}}$ and $\mathit{s^{i}_{2}}$ be the sets of $K$ nearest neighbors of $\mathbf{e_{i}}$ in two embedding spaces, respectively. We compute the average entity stability as $$\frac{1}{m} \sum_{i=1}^{m} \frac{\vert \mathit{s^{i}_{1}} \cap \mathit{s^{i}_{2}} \vert}{K}$$ which ranges between 0 and 1. A value of 1 indicates a perfect agreement between two embedding spaces while 0 indicates a complete disagreement.

      For entity-centric downstream tasks, one can run this experiment over a model $\mathbf{f}_{1}$ to first see if the retrieved sets of $K$ nearest neighbors to entities of interest fit their task domains. If not, one may want to try a different model $\mathbf{f}_{2}$ with a low entity stability relative to $\mathbf{f}_{1}$. This is because a model with high entity stability relative to $\mathbf{f}_{1}$ will be more likely to retrieve a set of entities similar to that of $\mathbf{f}_{1}$ and fail to fit task domains as well.
    \end{measure}
  \end{property}

  \begin{property}[Perturbation Robustness]\label{prop:pt_rb}
    Neural model performance has been found vulnerable to input perturbations. For example, state-of-the-art text-to-SQL models are shown to suffer from nuanced perturbations to database tables, natural language questions, and SQL queries~\cite{DBLP:journals/corr/chang2023drspider}. Such perturbations are designed to preserve semantics and can reveal a model's capacity to capture semantics. We hypothesize that preserving semantic similarities in the embedding space is key, especially, for downstream tasks such as retrieval, text-to-SQL and question answering. We therefore inspect the impact of input perturbations in the embedding space by measuring the robustness of column-level embeddings with respect to semantics-preserving perturbations.

    \begin{example*}
      Three database perturbations curated by ~\cite{DBLP:journals/corr/chang2023drspider} include \texttt{schema-synonym}, \texttt{schema-abbreviation}, and \texttt{column-} \texttt{equivalence}. \texttt{schema-synonym} and \texttt{schema-abbreviation} replace the name of a column with its synonym ("country" $\rightarrow$ "nation") and abbreviation ("CountryName" $\rightarrow$ "cntry\_name"), respectively. \texttt{column-equivalence} further perturbs both column names and contents, and may replace numerical columns with semantic-equivalent ones ("age" $\rightarrow$ "birthyear").
    \end{example*}

    \begin{measure}
      Given a set of original columns $\{ C_{i} \}_{i=1}^{n}$, we consider a set of perturbed variants $\{C'_{ij}\}_{j=1}^{m_{i}}$ for each $C_{i}$. The perturbations are semantics-preserving and can be at the schema level or data level or both. We compute the embedding cosine similarity of $(\textbf{E}(C_{i}), \textbf{E}(C'_{ij}))$ and average over all $m_{i}$ pairs for each $C_{i}$. We draw a distribution plot of average cosine similarity over $\{C_{i}\}_{i=1}^{n}$ across models and also report a single number of cosine similarity averaged over all $\sum_{i=1}^{n} m_{i}$ pairs for each model.
    \end{measure}
  \end{property}

  \begin{property}[Heterogeneous Context]\label{prop:context}
    Unlike coherent natural language sequences, tables are typically more heterogeneous comprising various types of data such as numeric, categorical, and datetime. As table embedding models mostly extend the architecture of language models and by default take context into consideration, it is less clear how much influence context has on embedding representations, especially for numeric data~\cite{DBLP:conf/kdd/HulsebosHBZSKDH19, DBLP:conf/sigmod/SuharaL0ZDCT22}. Without context (e.g., subject columns\footnote{The subject column of a table, if exists, contains the entities the table pertains to.} or neighboring columns), non-textual types of data, especially numerical columns, are typically hard to discriminate. Thus, it is important to understand the impact of context for many downstream tasks like semantic type prediction and relation extraction. In this property, we probe into the difference between contextual column embeddings and single-column embeddings for both textual and non-textual types of data.
    \begin{example*}
      Figure~\ref{fig:non_textual_data} shows a table from the SOTAB benchmark~\cite{DBLP:conf/semweb/KoriniPB22}. The table does not have a header and consists of both textual and non-textual data columns. Without context, \texttt{column} 4 is hard to interpret on its own, which could be percentages, prices or any metric numbers. However, the neighbor column to the right, namely \texttt{column} 5, which refers to the currency of Romania, can provide clues to the semantic meanings of \texttt{column} 4. In this context, it is more likely that \texttt{column} 4 contains price values.
    \end{example*}

    \begin{figure}[ht!]
      \centering
      \includegraphics[width=0.8\linewidth]{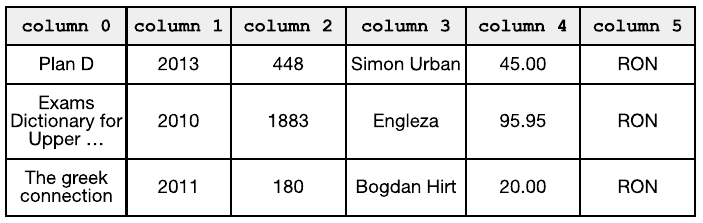}
      \caption{A table (without header) comprising textual and non-textual data columns.}
      \label{fig:non_textual_data}
    \end{figure}

    \begin{measure}
      To measure the effect of context, we consider four different input settings to get column embeddings as specified below. We compare embeddings of single columns with contextual column embeddings using their cosine similarity. 
      \begin{enumerate}[label=(\alph*), leftmargin=*]
        \item Only the column itself;
        \item Subject column as context (if not exist, use the first textual column from the left of a table as the proxy);
        \item Immediate neighboring columns on both sides as context; 
        \item The entire table as context.
      \end{enumerate}
    \end{measure}
  \end{property}

\section{Experiment Setup}
\subsection{Embedding Models}\label{sec:experiments_models}
  We consider well-established models and their variants that have been adopted for data management problems and open-sourced for public access. In particular, we select representative models from two categories: LMs and specialized table embedding models. Vanilla LMs are those designed for modeling natural language sequences and thus do not take into account the structure of tables or tabular data distributions. We include them in \sysname for comparison as many table embedding models share very similar architectures with weights initialized from LMs.

  \paragraph{Language Models} We take in \bert~\cite{DBLP:conf/naacl/DevlinCLT19}, \roberta~\cite{DBLP:journals/corr/liu19}, and \tfive~\cite{DBLP:journals/jmlr/RaffelSRLNMZLL20}. \bert is a pioneer transformer-based model that learns contextual representations from unlabeled text. \roberta builds on top of \bert and systematically studies the impact of key hyperparameters and training data size. Both models are go-to options for a wide range of NLP tasks and are bases for many tabular language models. \tfive is a representative of large language models whose largest variant has 11 billion parameters. We use base versions of all three models from the HuggingFace library~\cite{DBLP:conf/emnlp/WolfDSCDMCRLFDS20} in the experiments. 
  
  \begin{table}[ht!]
    \centering
    \caption{Overview of table embedding models and their design specifications (Column is abbreviated to Col.).}
    \label{tab:tabular_model_overview}
    \setlength{\tabcolsep}{2pt}
    \resizebox{\columnwidth}{!}{%
    \begin{tabular}{llll}
    \toprule
    \multicolumn{1}{c}{\textbf{Model}} &
        \multicolumn{1}{c}{\textbf{Input}} &
        \multicolumn{1}{c}{\textbf{Output Embedding}} &
        \multicolumn{1}{c}{\textbf{Downstream Task}} \\ \midrule
    \turl & Table + metadata & Entity / Col. / Col. pair & Table interpretation/augmentation \\
    \doduo & Table & Col. / Col. pair & Column type/relation prediction \\
    \tapas & NL question + table & Question / Table & Semantic parsing \\
    \tabert & NL question + table & Col. / Table & Semantic parsing \\
    \tapex & SQL query + table & Row / Table & Table Question Answering \\
    \taptap & Table & Row & Data augmentation/imputation \\ \bottomrule
    \end{tabular}%
    }
  \end{table}
  
  \paragraph{Table Embedding Models} We include \turl~\cite{DBLP:journals/pvldb/DengSL0020}, \doduo~\cite{DBLP:conf/sigmod/SuharaL0ZDCT22}, \tapas~\cite{DBLP:conf/acl/HerzigNMPE20}, \tabert~\cite{DBLP:conf/acl/YinNYR20}, \tapex~\cite{DBLP:conf/iclr/LiuCGZLCL22}, and \taptap~\cite{DBLP:journals/corr/abs-2305-09696}. \turl, \tapas, \tabert, \tapex, and \taptap first pretrain models over tables in an unsupervised manner by, for example, predicting masked column names or query execution results. The pretrained models are then fine-tuned for particular downstream tasks. We use pretrained models in the experiments as prescribed in our problem statement. \doduo directly fine-tunes a \bert-based model with labeled data from downstream tasks. See Table~\ref{tab:tabular_model_overview} for an overview of model specifications. The models we assess in experiments cover all levels of output embeddings, i.e., column, row, cell, and table embeddings.

\subsection{Datasets}
  We use both relational database tables and web tables for evaluation.

  \noindent\textbf{WikiTables.} The WikiTables~\cite{DBLP:conf/semweb/BhagavatulaND15} corpus contains 1.6M HTML tables of relational data extracted from Wikipedia pages. \turl preprocesses WikiTables and obtains an entity-rich dataset of 670,171 tables. We use the test partition released by TURL~\cite{turl_data}.

  \noindent\textbf{Spider.} Spider~\cite{DBLP:conf/emnlp/YuZYYWLMLYRZR18}, a widely-used semantic parsing and text-to-SQL dataset, includes 5,693 SQL queries over 200 databases across domains. We use the development set~\cite{spider_data} and run HyFD~\cite{DBLP:conf/sigmod/PapenbrockN16}, a functional dependency discovery algorithm, to create a dataset with annotated functional dependencies. To avoid mining a massive number of functional dependencies, we set the size of determinant to 1 and found 713 functional dependencies. We also collect an equal number of random pairs of columns without the relationship of functional dependencies for our experiments.
  
  \noindent\textbf{Dr.Spider.} Dr.Spider~\cite{DBLP:journals/corr/chang2023drspider} designs perturbations to databases, natural language questions, and SQL queries in Spider to test the robustness of text-to-SQL models. We take advantage of database perturbation tests in Dr.Spider~\cite{drspider_data} to evaluate the property of perturbation robustness.

  \noindent\textbf{NextiaJD.} Flores et al.~\cite{DBLP:conf/edbt/0002N021} collected 139 datasets from open repositories such as Kaggle and OpenML for predicting joinable columns. They also divided datasets into four testbeds based on dataset file size. For example, NextiaJD-XS includes datasets smaller than 1 MB while NextiaJD-L consists of datasets larger than 1 GB. Candidate pairs of columns are labeled with the join quality using a measure that takes account of both containment and cardinality proportion with empirically determined thresholds. For our evaluation, we use all pairs with join quality greater than 0.

  \noindent\textbf{SOTAB.} The Schema.org Table Annotation Benchmark~\cite{DBLP:conf/semweb/KoriniPB22} provides about 50,000 annotated tables collected from the WDC Schema.org Table Corpus for both column type and column property annotation tasks. We extract a subset that contains 5,000 tables for 20 semantic data types. The subset is balanced in terms of the number of non-textual and textual data types. Non-textual types include DATE, ISBN, POSTAL CODES, MONEY (monetary values), and QUANTITY (measurements as of weight etc.). We use this subset for measuring the property of Heterogeneous Context.

  Note that a dataset may not accommodate all the properties. For example, WikiTables does not have information of which two columns can be joined, so we do not measure the property of Join Relationship over WikiTables. On the other hand, properties such as Functional Dependencies and Heterogeneous Context require synthesized datasets for evaluation purposes. Table~\ref{tab:exp_obj_summary} summarizes the datasets and assessed models for each property. Also note that \turl, \tabert, and \taptap are excluded from certain experiments. This is because \turl is designed and implemented to output embeddings from entity-rich tables like those in WikiTables; \tabert yields only column embeddings after the fusion of the vertical attention mechanism; and \taptap encodes single rows independently using a text template serialization strategy and only gives row embeddings.

  \begin{table}[ht!]
    \caption{Overview of datasets and models for each property.}
    \label{tab:exp_obj_summary}
    \setlength{\tabcolsep}{2pt}
    \resizebox{\columnwidth}{!}{%
    \begin{tabular}{lll}
    \toprule
    \textbf{Property}                    & \textbf{Dataset}    & \textbf{Models in Scope} \\ \midrule
    Row order insignificance    & WikiTables & Except \taptap \\
    Column order insignificance & WikiTables & All                    \\
    Join relationship           & NextiaJD   & Except \turl and \taptap          \\
    Functional dependencies     & Spider     & Except \turl, \tabert, and \taptap \\
    Sample fidelity             & WikiTables & Except \taptap \\
    Entity stability            & WikiTables & Except \tabert and \taptap \\
    Perturbation robustness     & Dr. Spider & Except \turl and \taptap \\
    Heterogeneous Context       & SOTAB      & Except \turl and \taptap \\ \bottomrule
    \end{tabular}%
    }
  \end{table}

\subsection{Implementation}
  In general, we follow the original papers and their implementations in our evaluation. However, there are subtleties where extra consideration is needed, such as aligning the input and output across models for fair comparison. We make (minimal) design decisions in our implementation as discussed below.

  \paragraph{Table Serialization} As Transformer-based models expect to take sequence inputs, a key input processing step is to serialize two-dimensional tabular data into flattened sequences of tokens. Table embedding models considered in this analysis generally follow two common types of serialization methods.
  \begin{enumerate}[leftmargin=15pt]
    \item Row-wise serialization. Tables are parsed by rows, which are further concatenated with optional insertions of special tokens as delimiters. \turl, \tapas, and \tabert fall under this category despite the difference that \tapas uses dedicated positional embeddings to indicate the row and column in which a token appears while \tabert explicitly adds [SEP] tokens to mark boundaries of cells in the sequences.
    \item Column-wise serialization. Alternatively, tables can be serialized by column. For \doduo, [CLS] tokens (as many as the number of columns) are inserted to separate values from different columns and are effectively used as column representations.
  \end{enumerate}  

  For each table embedding model, we adopt the serialization method as proposed in the original papers. Since vanilla language models do not have a default serialization method for tabular data, we experimentally apply row/column-wise serialization as applicable. In practice, models also enforce a length limit to token sequences (e.g. 512 is a common maximum). To ensure that all models take in (almost) the same inputs regardless of serialization methods, we keep all the columns for each table, if possible, and preserve as many rows as the length limit permits. We use binary search to find the maximum number of rows that can fit into the input limit. 

  \paragraph{Embedding Retrieval} We use the embeddings provided by a model, if they are available. However, due to designs for particular downstream tasks, a model may not readily expose certain levels of embeddings needed for measuring a property. For instance, \tapas does not give row or column embeddings out of the box. We circumvent this obstacle by observing that all the models can output token-level embeddings and some table embedding models have additional mask embeddings or positional embeddings that indicate to which row and column a token belongs. Therefore, we can aggregate token embeddings (by averaging them for example) to embeddings on a level (e.g. row or column) as needed. In particular, we take advantage of different serialization methods and use special tokens to retrieve row or column or table embeddings. As to cell embeddings needed for the property Functional Dependencies and entity embeddings as needed for the Entity Stability property, we keep track of token positions in the table and aggregate them accordingly. We take this alternative since inserting special tokens for each cell quickly uses up the input limit.
    
  The practice of inserting special tokens and aggregating lower level of embeddings is common in the literature~\cite{DBLP:conf/acl/YinNYR20, DBLP:journals/pvldb/DengSL0020, DBLP:journals/pvldb/0001LSDT20, DBLP:conf/sigmod/SuharaL0ZDCT22, DBLP:journals/corr/cong2023pylon}. As noted in our problem statement (Section~\ref{sec:methodol_ps}), we consider pretrained models in \sysname, thereby not fine-tuning any model for downstream tasks. We illustrate in Section~\ref{sec:conn} that the characterization of pretrained models remain effective for anticipating behaviors of finetuned models on various downstream tasks.

\section{Results}
  In this section, we present the experiment results and our analysis and describe the characteristics of models over the eight properties.

\subsection{Row Order Insignificance}
  \begin{figure}[!t]
    \centering
    \minipage{0.49\columnwidth}
      \includegraphics[width=\textwidth]{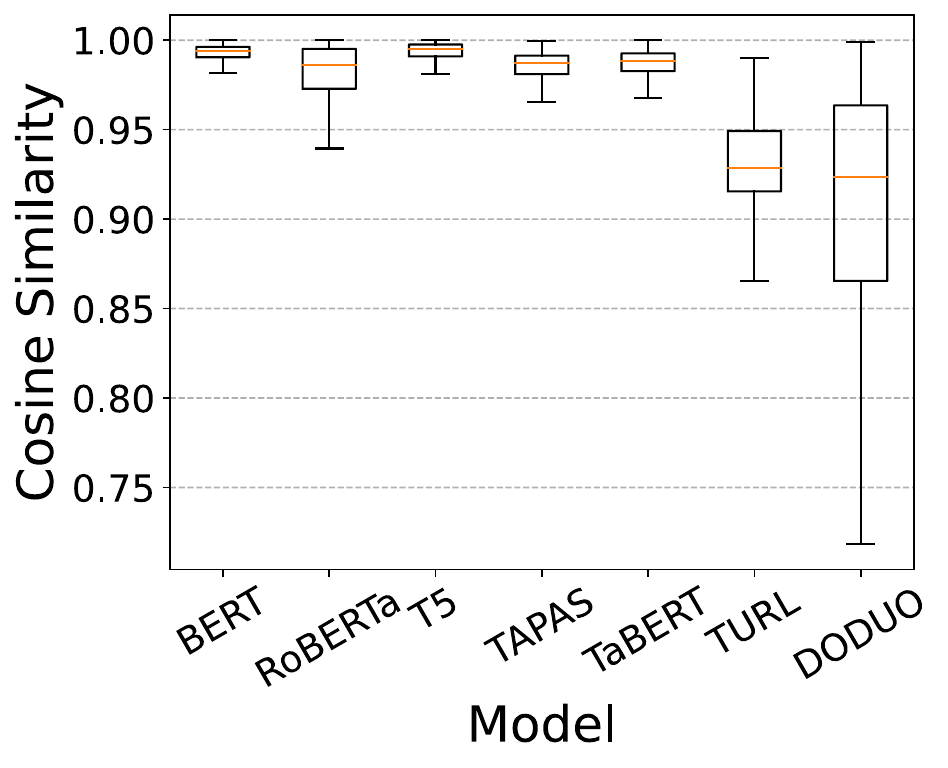}
    \endminipage\hfill
    \minipage{0.49\columnwidth}
      \includegraphics[width=\textwidth]{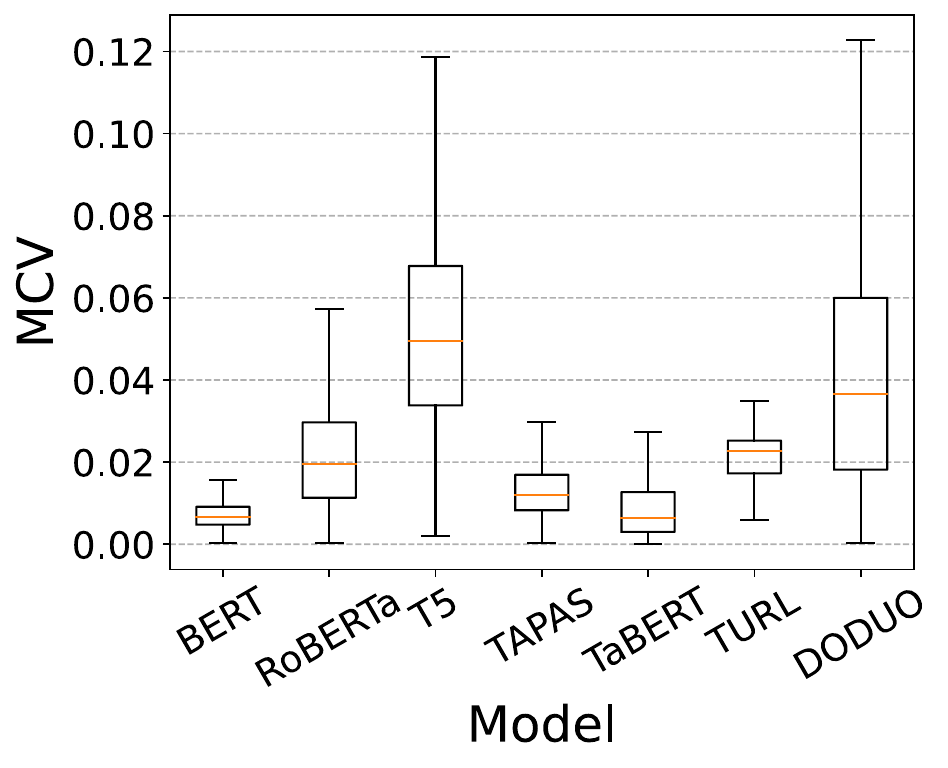}
    \endminipage \\
    \minipage{0.49\columnwidth}
      \includegraphics[width=\textwidth]{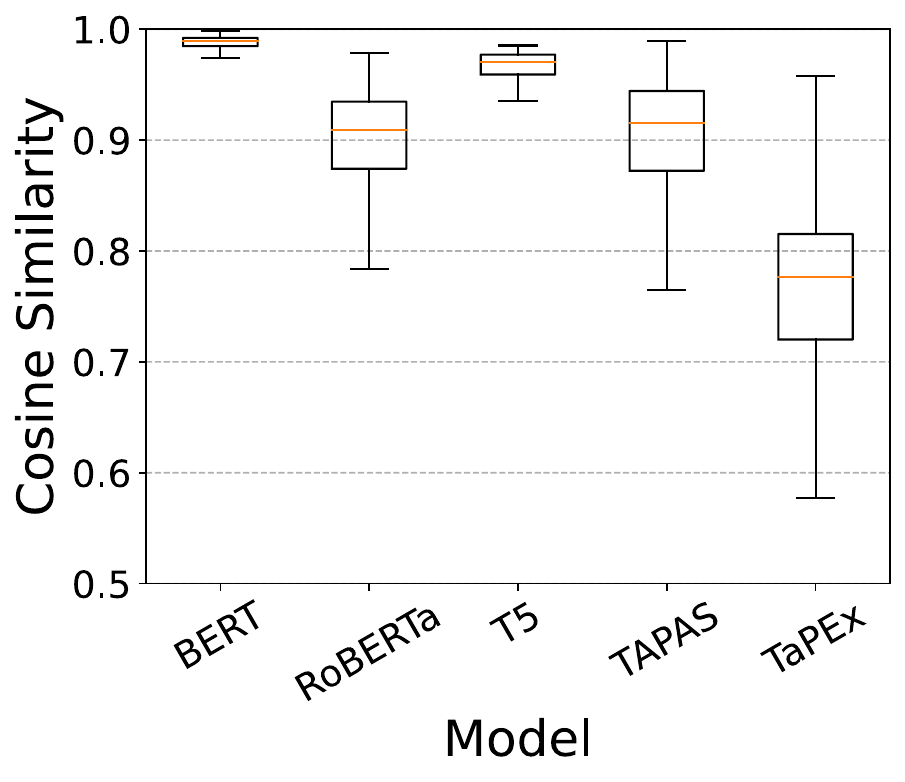}
    \endminipage\hfill
    \minipage{0.49\columnwidth}
      \includegraphics[width=\textwidth]{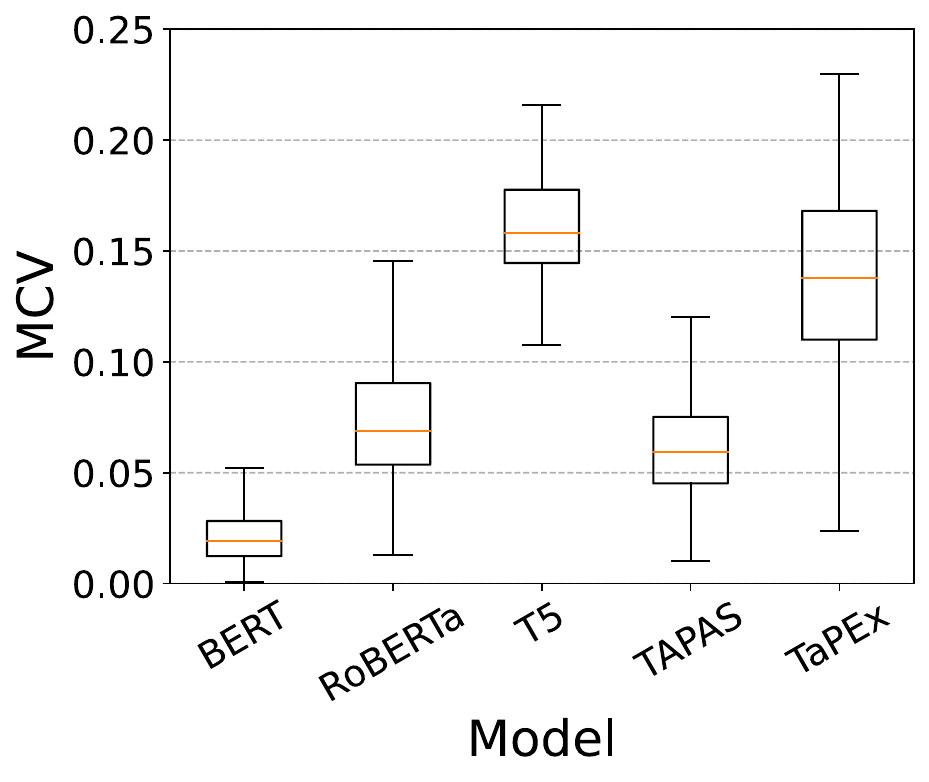}
    \endminipage \\
    \minipage{0.49\columnwidth}
      \includegraphics[width=\textwidth]{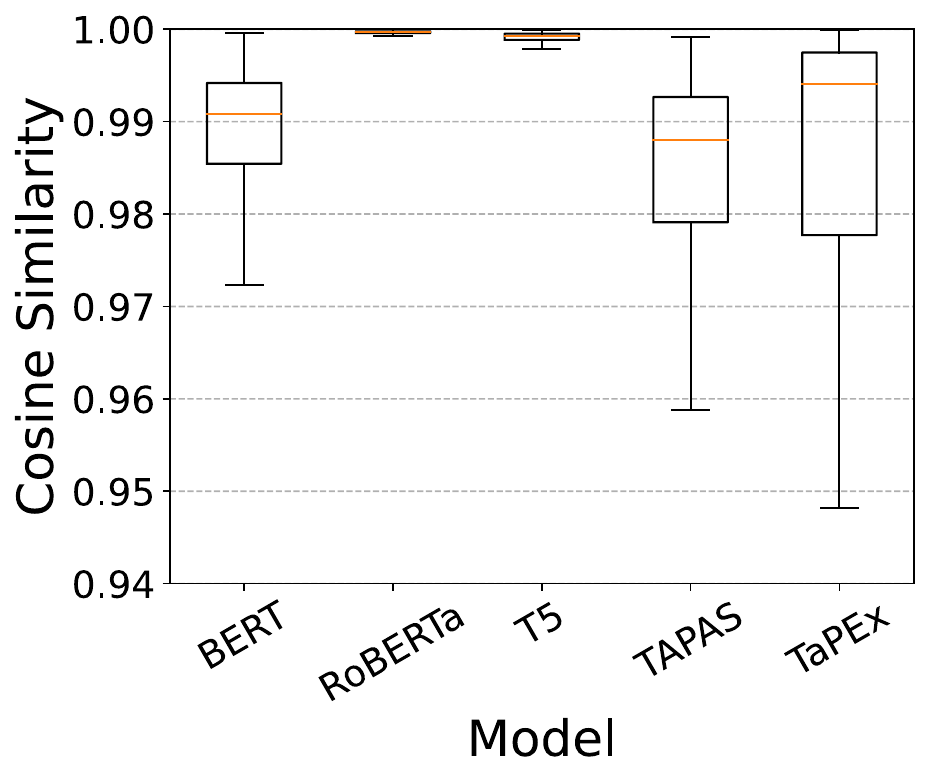}
    \endminipage\hfill
    \minipage{0.49\columnwidth}
      \includegraphics[width=\textwidth]{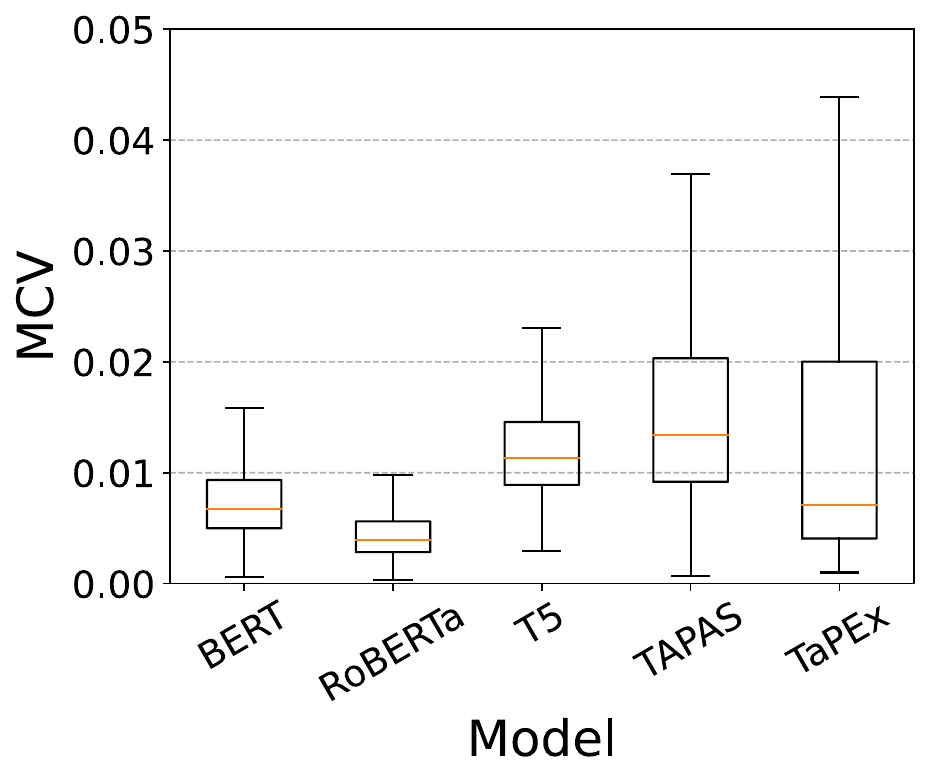}
    \endminipage
    \caption{Cosine similarity and MCV distributions of column (top), row (middle), and table (bottom) embeddings from row shuffling. Across three levels of embeddings, table embedding models exhibit comparably lower cosine similarity while both language and table embedding models may exhibit high MCV.}
    \label{fig:row_insig}
  \end{figure}

  \begin{figure}[!t]
    \centering
    \minipage{0.49\columnwidth}
      \includegraphics[width=\textwidth]{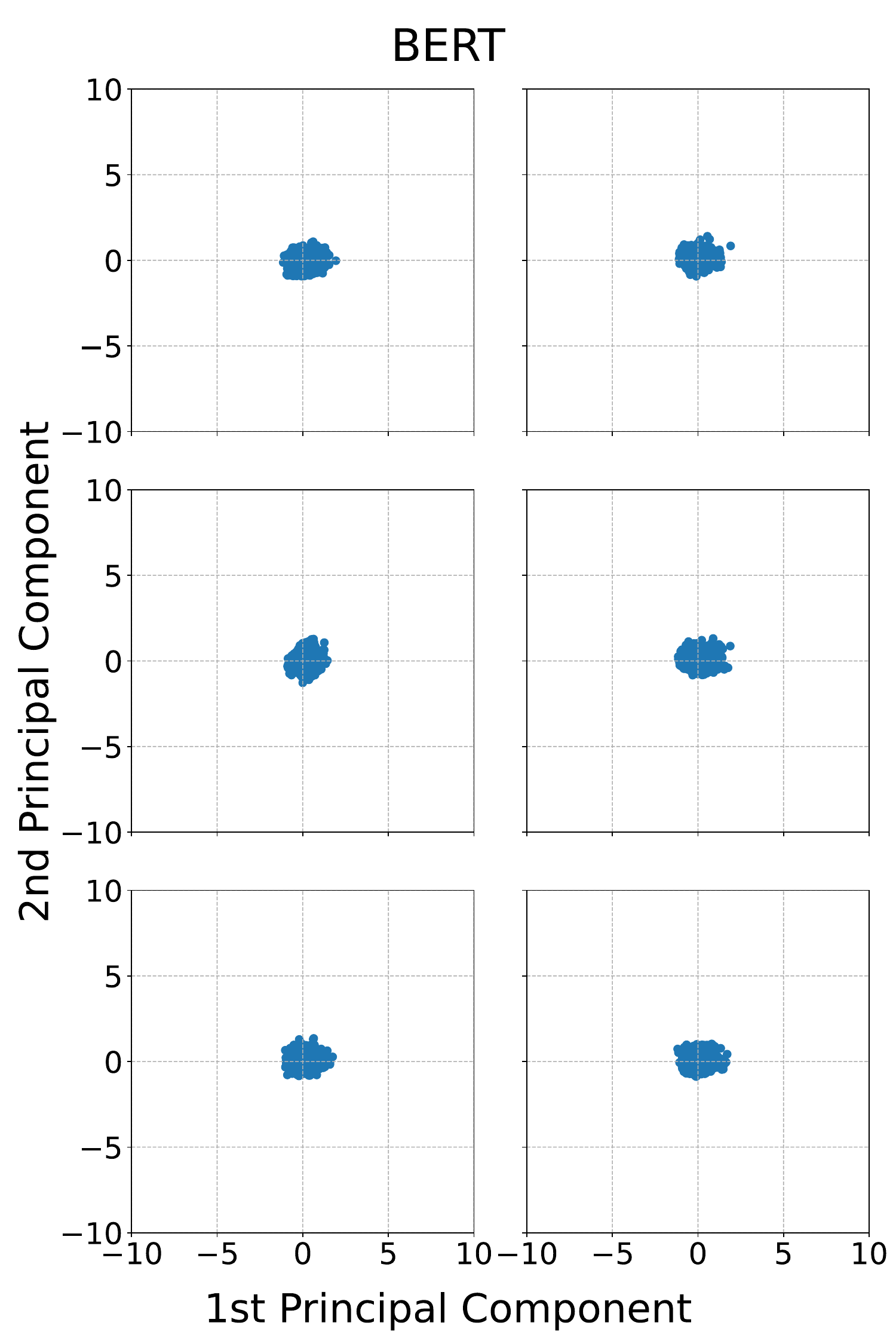}
    \endminipage\hfill
    \minipage{0.49\columnwidth}
      \includegraphics[width=\textwidth]{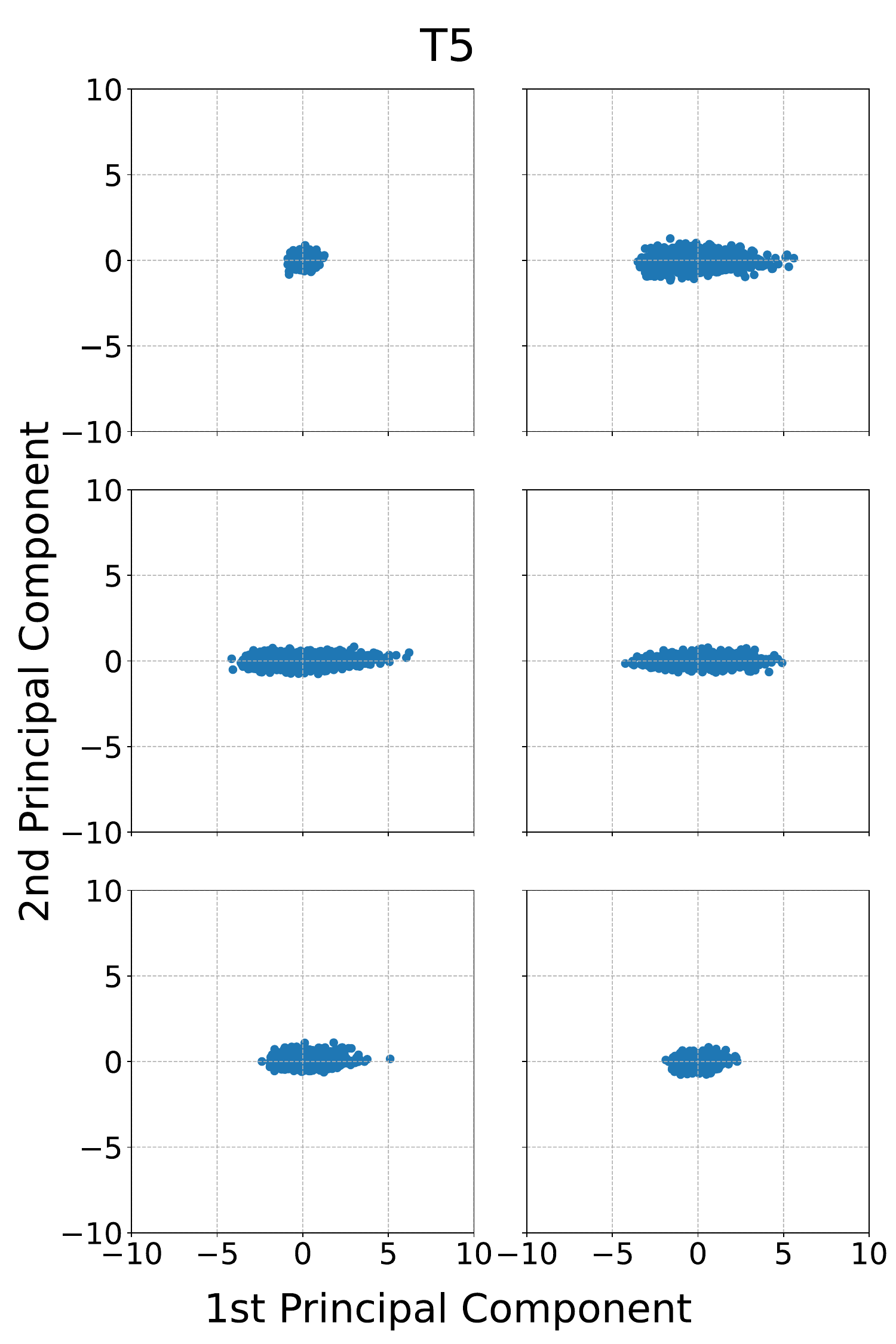}
    \endminipage
    \caption{PCA visualization of high-dimensional column embeddings from a table of six columns, for \bert and \tfive. Each subplot draws 6!=720 row-wise permutation variants of a column. While \bert embeddings are centered around the origin with some variation, the \tfive embeddings are more stretched along the horizontal axis, resulting in the relatively high cosine similarity as well as high MCV value.}
    \label{fig:row_insig_pca}
  \end{figure}

  We calculate cosine similarity and MCV (as defined in Measure~\ref{measure:row_order}) measures over column, row, and table embeddings and plot their distributions in Figure~\ref{fig:row_insig}. Overall, table embedding models exhibit comparably lower cosine similarity while both language and table embedding models may exhibit high MCV.

  On the top row of Figure~\ref{fig:row_insig}, column embeddings of five models \bert, \roberta, \tfive, \tapas, and \tabert show strong evidence of being robust to row order shuffling in terms of cosine similarity. In particular, the first quartile (Q1) of these models is above 0.97 and the minimum (Q1 - 1.5 $\times$ interquartile range) is above 0.95 except \roberta. \turl follows with Q1 above 0.92 and the minimum above 0.86. \doduo exhibits the largest spread with the minimum below 0.75 while the median is over 0.91. This implies that \doduo is relatively sensitive towards row order, given that the content of these rows is not altered. We illustrate in Section~\ref{sec:conn} how \doduo's sensitivity to row order shuffling translates to unstable predictions in a downstream task for which \doduo is proposed.

  The MCV measure (the lower the better) indicates the variability of different populations (i.e., embedding distributions given by different models), especially when they have different means. It is notable that \tfive has the largest third quartile (Q3) and second largest maximum (Q3 + 1.5 $\times$ interquartile range) while \tfive embeddings have high cosine similarities. We hypothesize that this is because \tfive embeddings are more dispersed in a specific direction in high-dimensional space compared to models with low MCVs such as \bert. We verify this by visualizing the PCA projections of embeddings in two-dimensional space. For demonstration purposes, we use a table in which \tfive embeddings of three columns yield high MCV scores (larger than 0.08, which is higher than Q3). Correspondingly, the projections of \tfive embeddings of these three columns (top-right, middle left and middle right) are indeed more stretched along a specific direction than those of \bert as demonstrated in Figure~\ref{fig:row_insig_pca}.

  With regard to row embeddings (the middle row of Figure~\ref{fig:row_insig}), it is noticeable that \bert obtains high cosine similarity with the minimum above 0.95 and low MCV with Q3 below 0.03. On the other hand, row embeddings of \roberta, \tfive, \tapas and \tapex appear to vary more in MCV than their column embeddings.
  As seen in the bottom row of Figure~\ref{fig:row_insig}, unlike column and row embeddings, table embeddings of assessed models manifest exceptionally high cosine similarity and low MCV. Precisely, the minimum of the cosine similarity of each model is above 0.94 and the range of MCV is also 5$\times$ smaller than that of row embeddings.

  These findings also underline the importance of combining the cosine similarity and MCV for measuring row order insignificance, as a single measure would give a limited perspective.

\subsection{Column Order Insignificance}
  \begin{figure}[!t]
    \centering
    \minipage{0.49\columnwidth}
      \includegraphics[width=\textwidth]{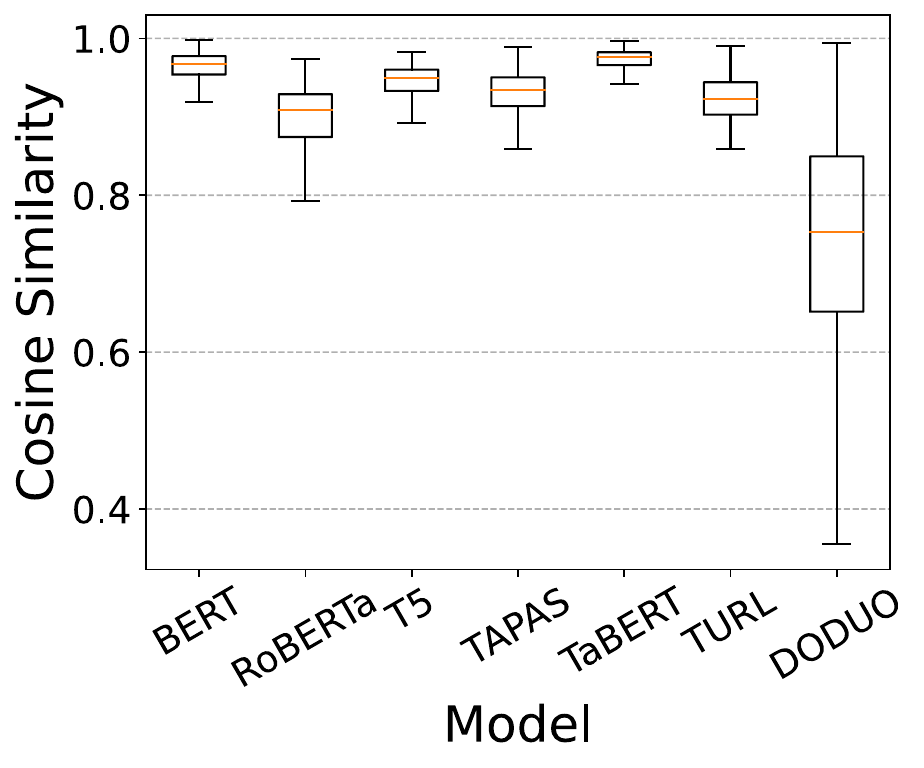}
    \endminipage\hfill
    \minipage{0.49\columnwidth}
      \includegraphics[width=\textwidth]{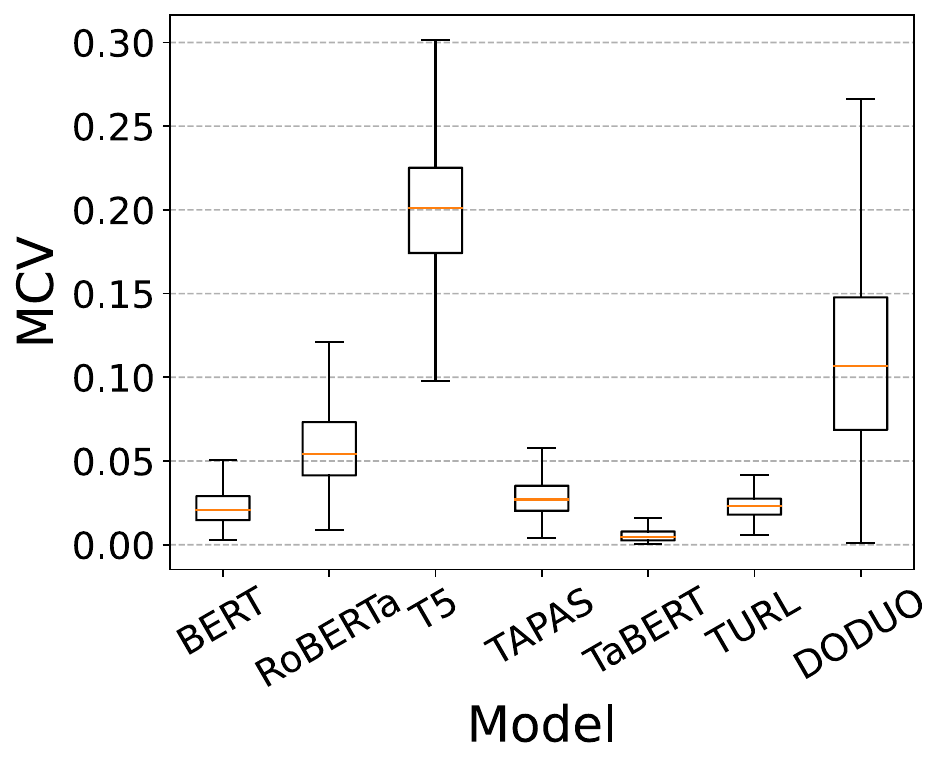}
    \endminipage \\
    \minipage{0.49\columnwidth}
      \includegraphics[width=\textwidth]{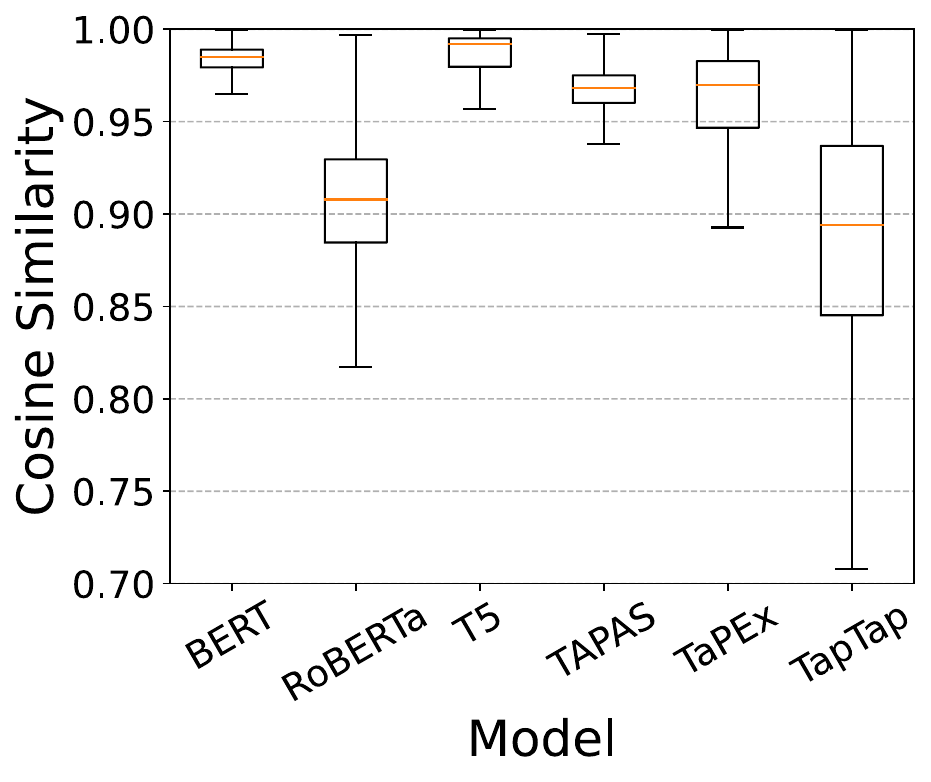}
    \endminipage\hfill
    \minipage{0.49\columnwidth}
      \includegraphics[width=\textwidth]{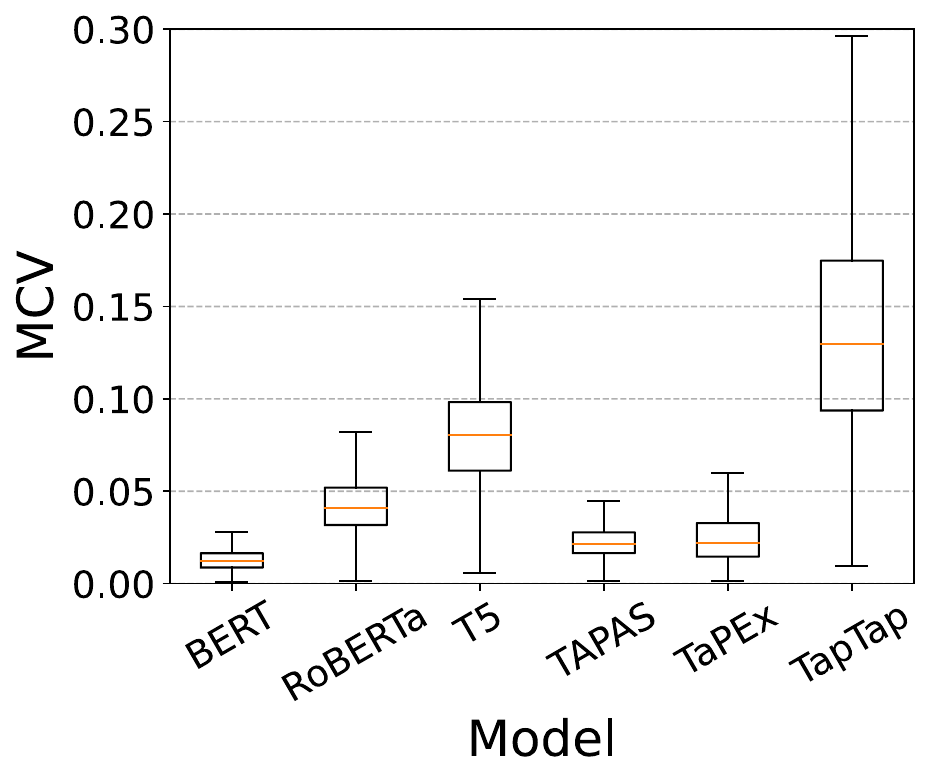}
    \endminipage
    \caption{Cosine similarity and MCV distributions of column (top) and row (bottom) embeddings from column shuffling. Both column and row embeddings manifest similar patterns as in row shuffling.}
    \label{fig:col_insig}
  \end{figure}

  \begin{figure}[!t]
    \centering
    \minipage{0.49\columnwidth}
      \includegraphics[width=\textwidth]{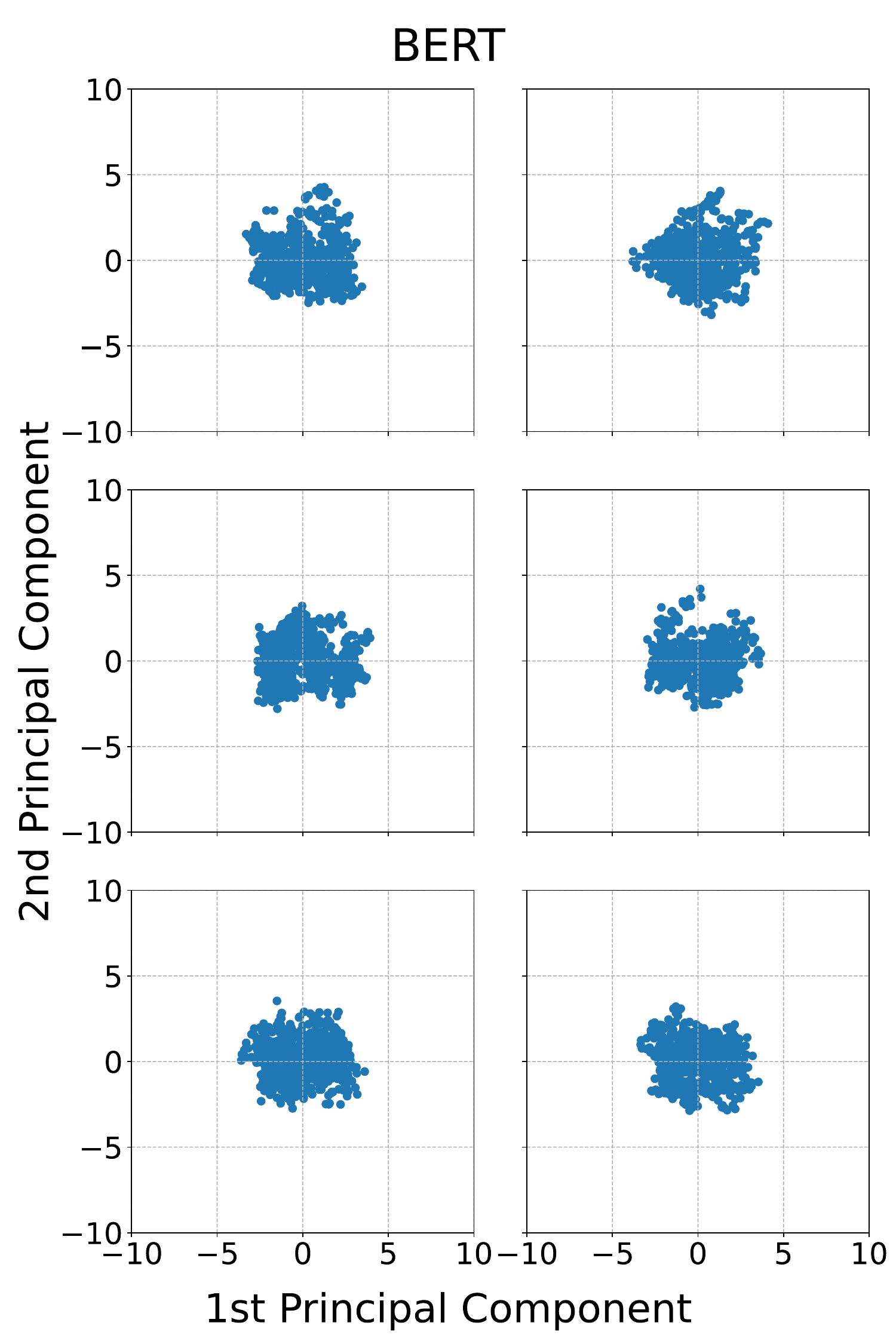}
    \endminipage\hfill
    \minipage{0.49\columnwidth}
      \includegraphics[width=\textwidth]{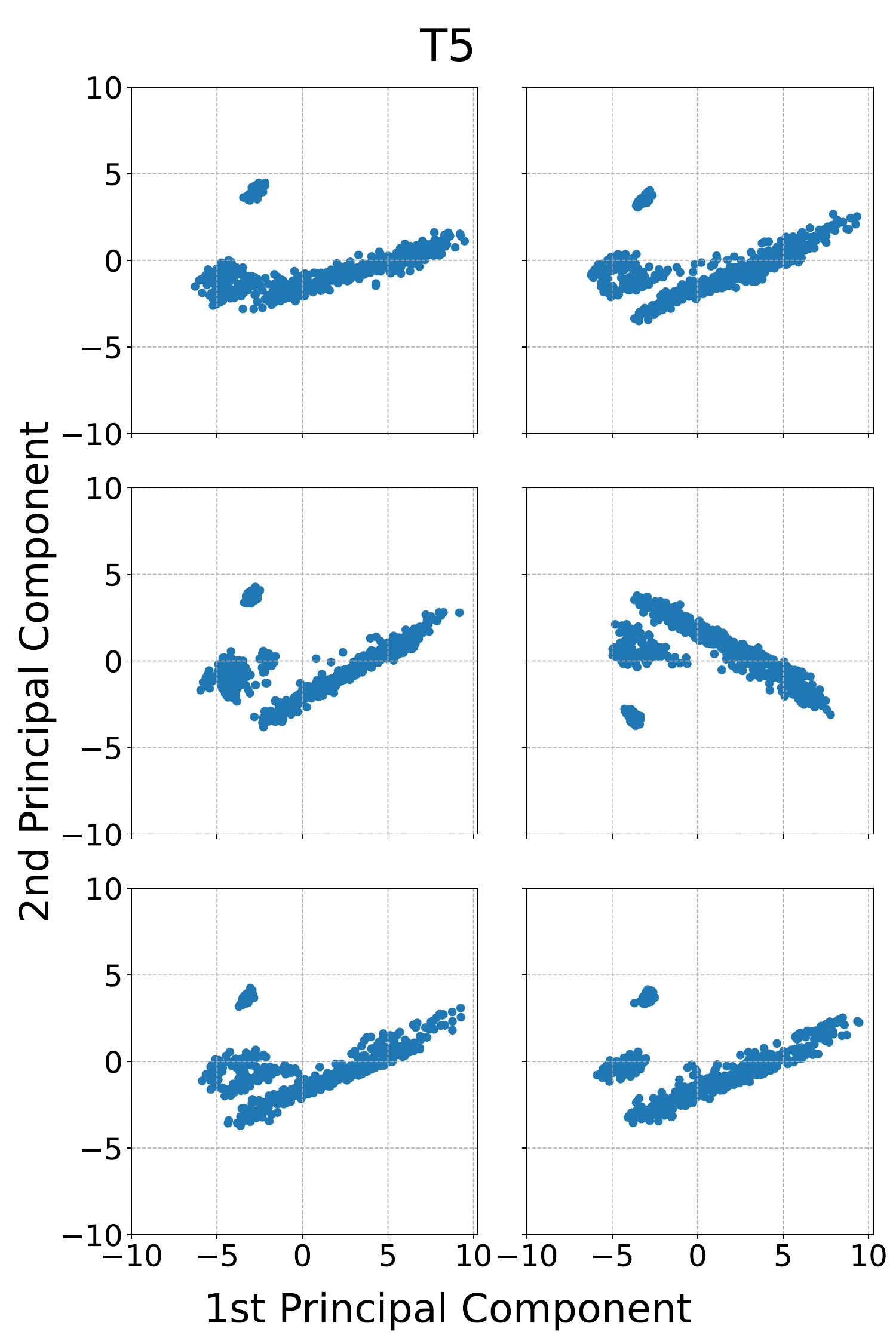}
    \endminipage
    \caption{PCA visualization of high-dimensional column embeddings from the same table as used in Figure~\ref{fig:row_insig_pca}. Each subplot draws 6!=720 variants of a column from column order shuffling. The embeddings exhibit similar patterns as in row order shuffling but show larger spread across all columns.}
    \label{fig:col_insig_pca}
  \end{figure}

  The results of column shuffling follow a similar trend as that of row shuffling. Nevertheless, column shuffling appears to cause more variations in all three levels of embeddings for both cosine similarity and MCV measures. In the interest of space, we only show column and row embeddings in Figure~\ref{fig:col_insig}.
  
  Considering, for example, the column embeddings in Figure~\ref{fig:col_insig}, the median cosine similarity of \roberta embeddings drops by more than $5\%$ and the same statistic of \doduo embeddings drops by more than $15\%$. The median MCV of both \roberta and \tfive also increases by four times. To verify such large variations, we again visualize the PCA projections of \tfive embeddings in Figure~\ref{fig:col_insig_pca} for the same table as used in Figure~\ref{fig:row_insig_pca}. This figure confirms that the first principal component of \tfive embeddings manifests larger spread, and illustrates the spread along the horizontal axis across all columns (instead of merely 3, as when rows are shuffled) indicating a higher sensitivity to column order than row order.

\subsection{Join Relationship}
  \begin{table}[!ht]
    \caption{Spearman coefficients between a value overlap measure and embedding cosine similarity on the NextiaJD-XS dataset. Multiset Jaccard is most positively correlated to embedding cosine similarity across all models. All coefficient numbers are statistically significant ($\text{p-value} < 0.01$).}
    \label{tab:join_relationship_xs}
    \setlength{\tabcolsep}{2pt}
    \setlength{\extrarowheight}{-2pt}
    \resizebox{\columnwidth}{!}{%
      \begin{tabular}{lcccccc}
      \toprule
      \multicolumn{1}{l}{} & \textbf{BERT}  & \textbf{RoBERTa} & \textbf{T5}    & \textbf{TAPAS} & \textbf{TaBERT} & \textbf{DODUO} \\ \midrule
      Containment          & 0.241 & 0.412   & 0.649 & 0.438 & 0.506  & 0.438 \\
      Jaccard              & 0.288 & 0.339   & 0.563 & 0.368 & 0.553  & 0.441 \\
      Multiset Jaccard     & 0.670 & 0.512   & 0.647 & 0.655 & 0.721  & 0.696 \\ \bottomrule
      \end{tabular}%
    }
  \end{table}  

\begin{figure}[!t]
    \centering
    \includegraphics[width=\columnwidth]{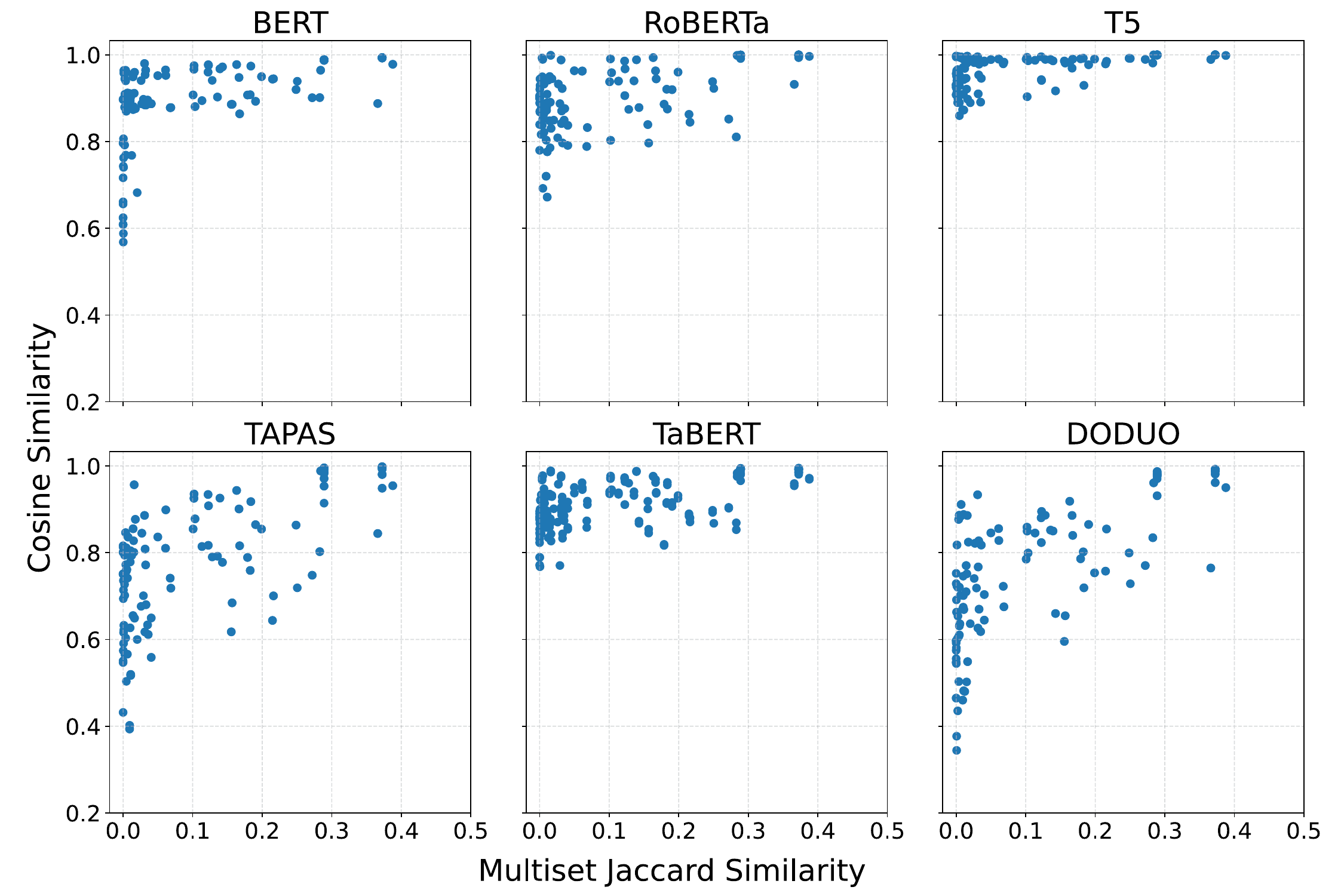}
    \caption{Scatter plots of embedding cosine similarity vs. multiset Jaccard similarity derived from pairs of joinable columns in the NextiaJD XS dataset, which illustrate a positive correlation between the two measures.}
    \label{fig:join_multiset_jaccard}
  \end{figure}

  Table~\ref{tab:join_relationship_xs} presents the Spearman coefficients between a value overlap measure and embedding cosine similarity over joinable pairs of columns from the NextiaJD-XS dataset. We find that, among the considered value overlap measures (containment, Jaccard, and multiset Jaccard), multiset Jaccard similarity is most positively correlated with embedding cosine similarity. For all models, the coefficient value between multiset Jaccard and embedding cosine similarity is above 0.5, which indicates a moderate positive correlation (\tabert has a coefficient value of 0.72 which indicates a high positive correlation), and is significantly higher than that of the other two measures ($0.08 - 0.43 \uparrow$). This difference can be attributed to the fact that containment and Jaccard similarity do not take duplicate values into account while we use all values for embedding inference. In Figure~\ref{fig:join_multiset_jaccard}, we also show scatter plots of embedding cosine similarity versus multiset Jaccard over pairs of joinable columns from NextiaJD-XS for each model, which demonstrates the moderate positive correlation between the two variables. Note that the maximum possible value of multiset Jaccard similarity is 0.5.

  Both syntactic and semantic approaches have been employed for data discovery~\cite{DBLP:journals/pvldb/NargesianZPM18, DBLP:conf/icde/BogatuFP020, DBLP:journals/corr/cong2023pylon}. It is valuable to be aware of what syntactic measure is highly correlated with a semantic measure based on embeddings so that one can ensemble less correlated syntactic and embedding measures if they want to find more diverse candidates. For instance, consider the task of join discovery over NextiaJD-XS. Based on  Table~\ref{tab:join_relationship_xs}, it is recommended to use containment as the syntactic similarity measure when \bert embeddings are used to measure semantic similarities because these two measures show the least correlation. Similarly, it is recommended to use the Jaccard similarity when \tapas embeddings are used.

\subsection{Functional Dependencies}
  \begin{table}[!h]
    \caption{Average group-wise variances of embedding translations over columns with and without functional dependencies across five models. Only \tapas yields $\overline{S^2}_{FD} < \overline{S^2}_{\neg{FD}}$ with $\overline{S^2}_{FD}$ close to 0, while language models and other table embedding models do not follow this pattern.}
    \label{tab:fd}
    \setlength{\tabcolsep}{2pt}
    \resizebox{\columnwidth}{!}{%
    \begin{tabular}{lccccc}
    \toprule
    \multicolumn{1}{c}{}      & \textbf{BERT}   & \textbf{RoBERTa} & \textbf{T5}     & \textbf{TAPAS}  & \textbf{DODUO}   \\ \midrule
    Columns w/ FD     & 0.87 & 0.39  & 1.80 & 0.88 & 83.34  \\ 
    Columns w/o FD & 0.78 & 0.34  & 1.13 & 1.12 & 229.77 \\ \bottomrule
    \end{tabular}%
  }
  \end{table}

  \begin{figure*}[!ht]
    \centering
    \minipage{0.2\textwidth}
      \includegraphics[width=\textwidth]{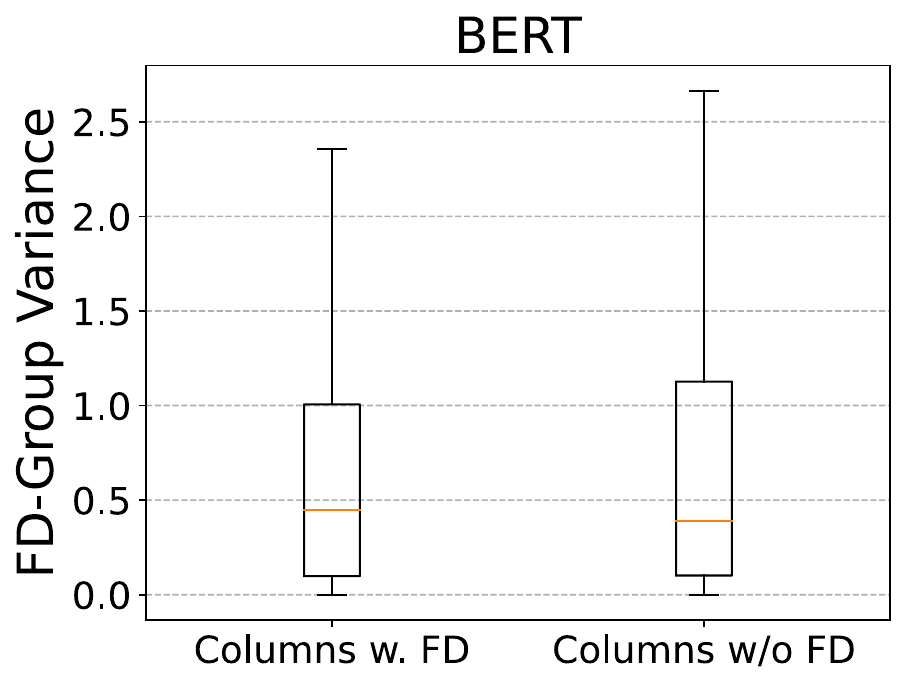}
    \endminipage\hfill
    \minipage{0.2\textwidth}
      \includegraphics[width=\textwidth]{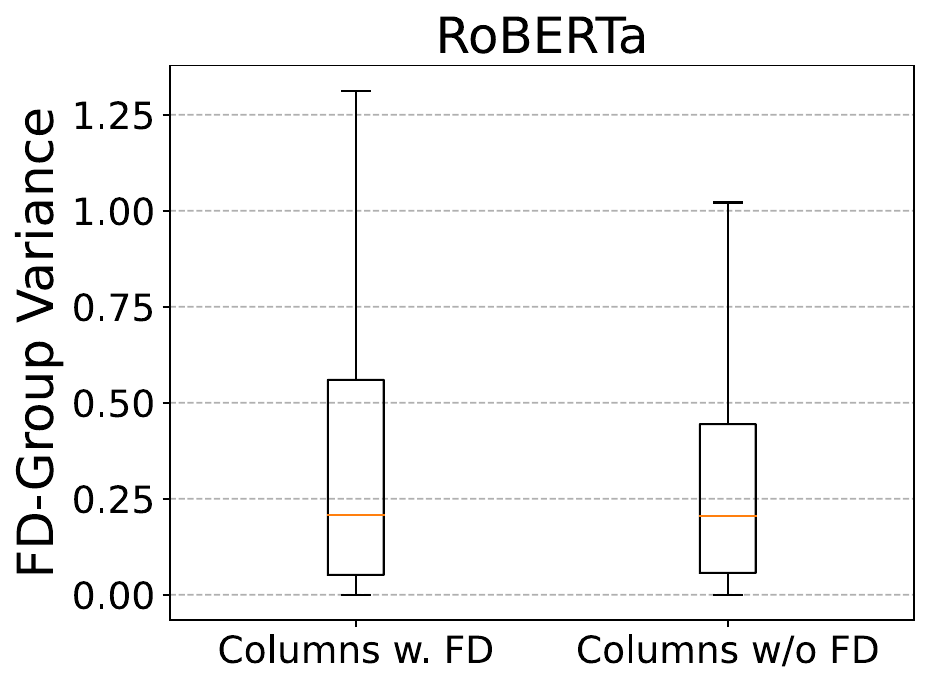}
    \endminipage\hfill
    \minipage{0.19\textwidth}
      \includegraphics[width=\textwidth]{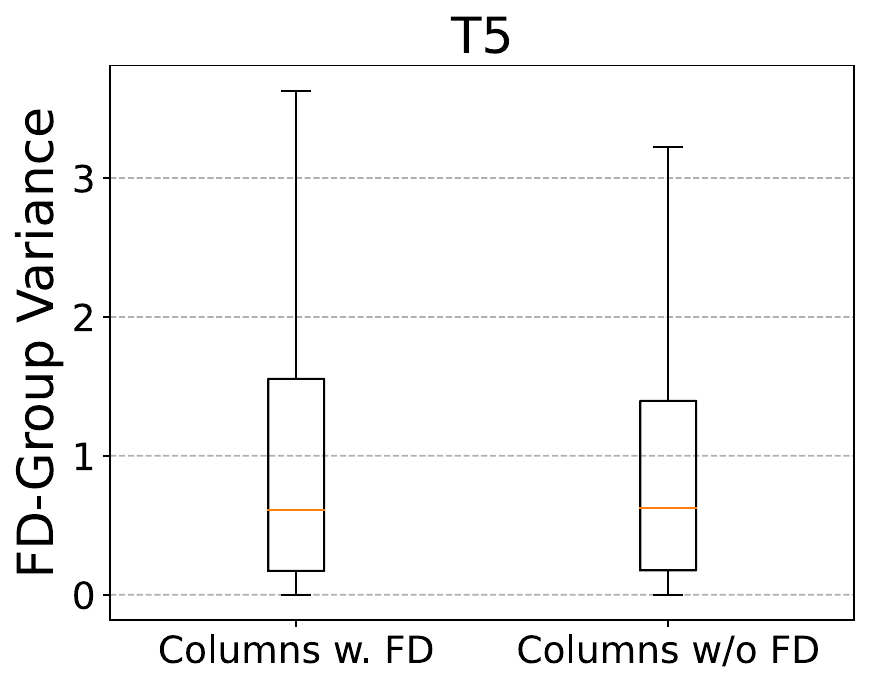}
    \endminipage\hfill
    \minipage{0.19\textwidth}
      \includegraphics[width=\textwidth]{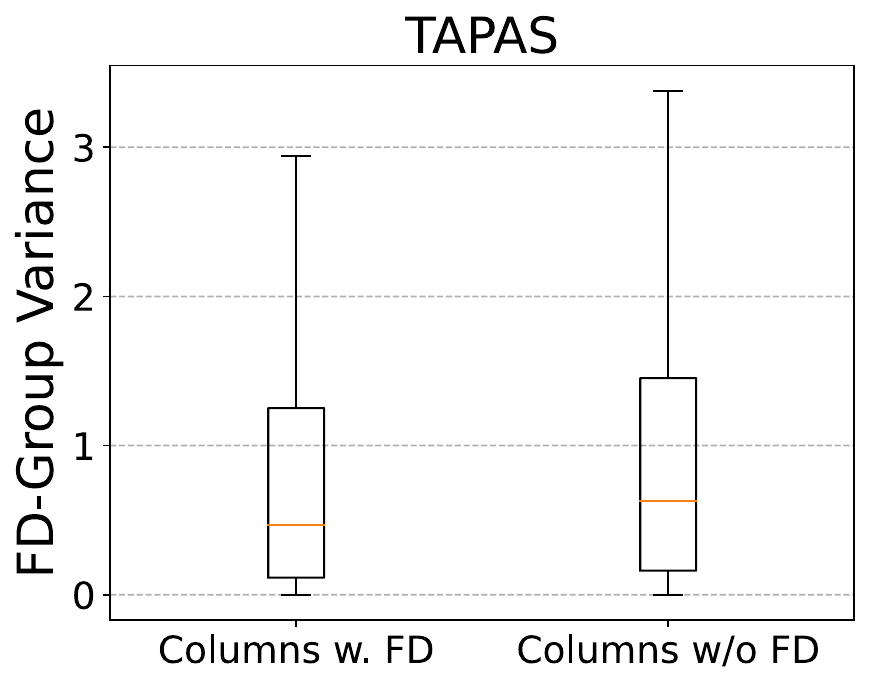}
    \endminipage\hfill
    \minipage{0.2\textwidth}
      \includegraphics[width=\textwidth]{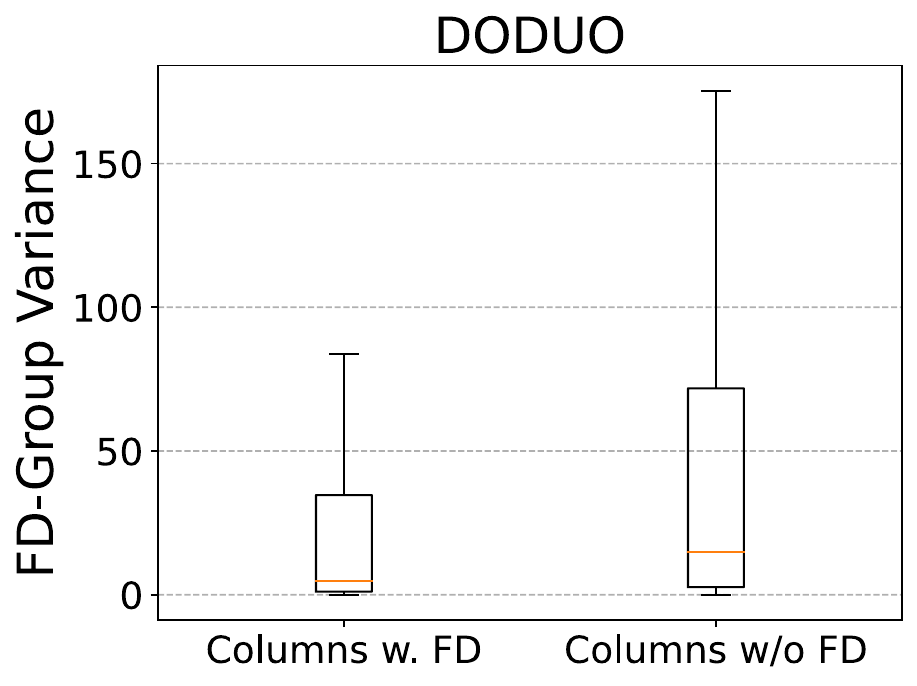}
    \endminipage
    \caption{Distributions of the group-wise variances over embedding translations across column pairs with and without the relationship of functional dependencies. None of the models show clear separation between the two variance distributions.}
    \label{fig:fd_boxplots}
  \end{figure*}


  Table~\ref{tab:fd} displays the average variance of the L2 norm of translation embeddings over column pairs, comparing those with and without functional dependencies. Vanilla language models exhibit no significant reduction in variance for columns with functional dependencies, as expected, given their lack of consideration for table structure during pretraining. In contrast, table embedding models, including \doduo and \tapas, show variance patterns contrary to vanilla language models, though the average variance of \doduo is not close to 0. Despite \tapas aligning with expected patterns, Figure~\ref{fig:fd_boxplots} reveals that none of the models distinctly separate variance distributions for column pairs with and without functional dependencies. This lack of clear separation provides evidence that none of the models effectively capture the relationship of functional dependencies in their representations.

\subsection{Sample Fidelity}
  \begin{figure}[!ht]
    \centering
    \minipage{0.33\columnwidth}
      \includegraphics[width=\textwidth]{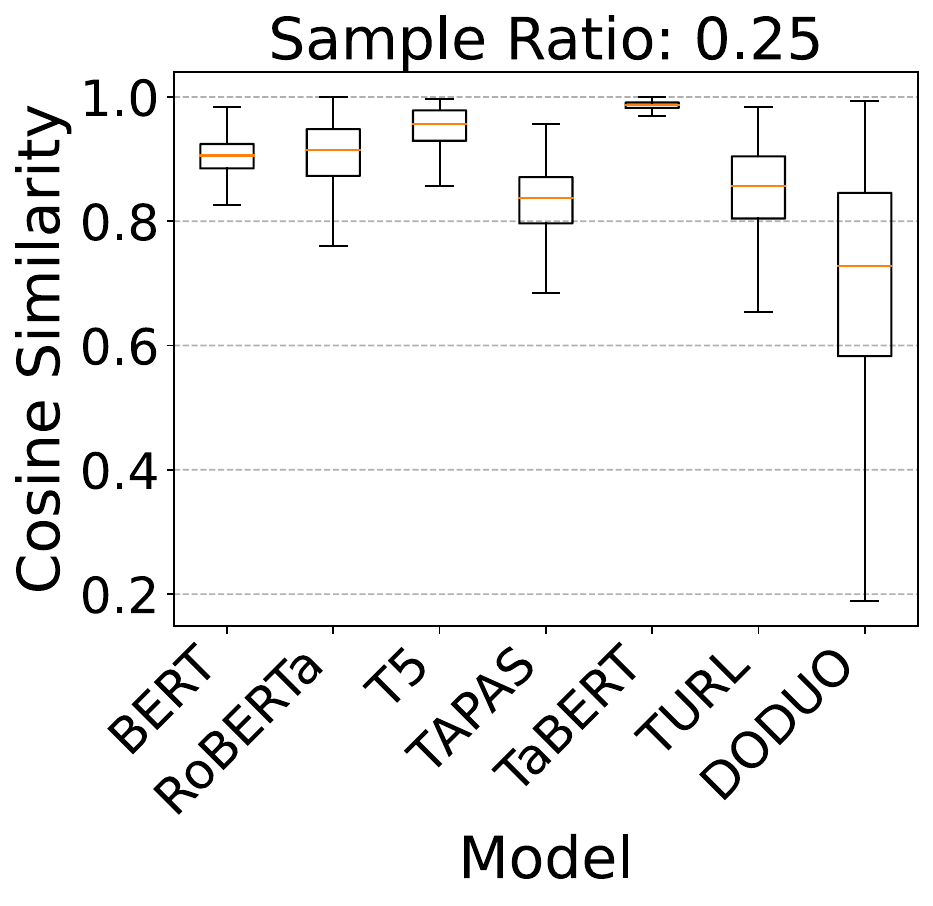}
    \endminipage\hfill
    \minipage{0.33\columnwidth}
      \includegraphics[width=\textwidth]{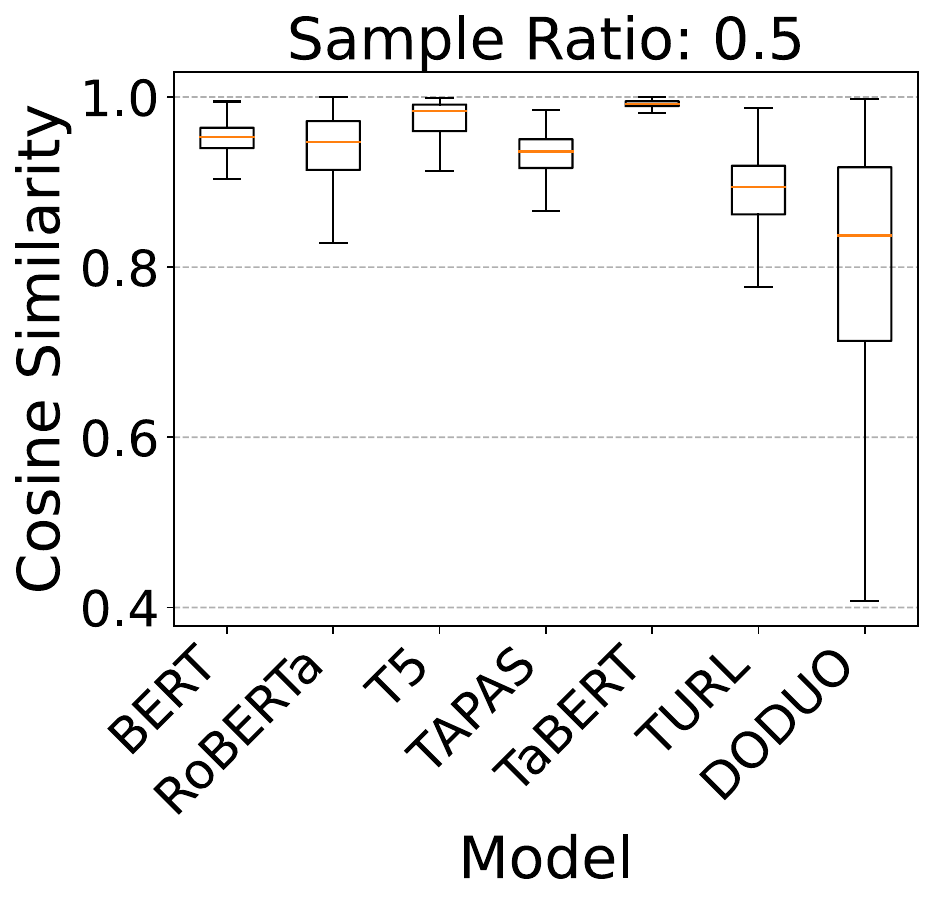}
    \endminipage\hfill
    \minipage{0.33\columnwidth}
      \includegraphics[width=\textwidth]{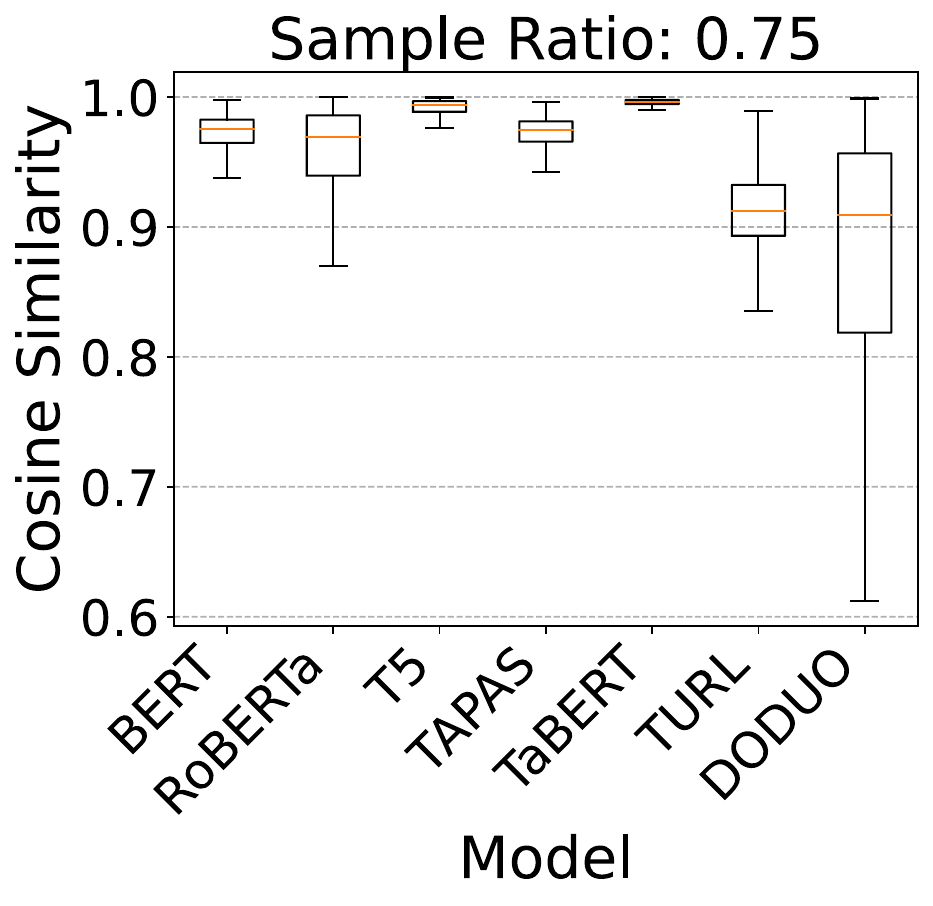}
    \endminipage
    \caption{Distributions of sample fidelity of column embeddings under three sample ratios. Overall, vanilla LMs exhibit higher sample fidelity compared to table embedding models.}
    \label{fig:sample_fidelity}
  \end{figure}

  Figure~\ref{fig:sample_fidelity} depicts sample fidelity distributions of models across various sample ratios. As the ratio increases, sample embeddings tend to align more closely with those from full values in terms of cosine similarity, evident in the ascending quartile values of embedding cosine similarity in the box plots.

  Vanilla language models consistently show high sample fidelity, reaching a median over 0.9 at a 0.25 sample ratio and exceeding 0.95 at a 0.75 ratio. Notably, \tfive demonstrates strong robustness to sampling, with over 75\% of tested pairs having cosine similarity surpassing 0.95 when half of the values are sampled. Table embedding models, excluding \tabert, exhibit larger distribution spreads, particularly at a 0.25 sample ratio. \tabert stands out as the most sample-robust model, consistently maintaining cosine similarity over 0.95 across all sample ratios. This robustness stems from \tabert's internal practice of always considering the first three rows~\cite{tabert_config}, increasing the likelihood of overlapping or identical inputs despite sampling. While \tapas emerges as the next sample-robust model, achieving high fidelity comparable to vanilla language models at a 0.5 sample ratio, \doduo lags behind and proves more sensitive to sampling across all ratios, consistent with the results of row and column shuffling.

\subsection{Entity Stability}
  \begin{figure}[!ht]
    \centering
    \minipage{0.305\columnwidth}
      \includegraphics[width=\textwidth]{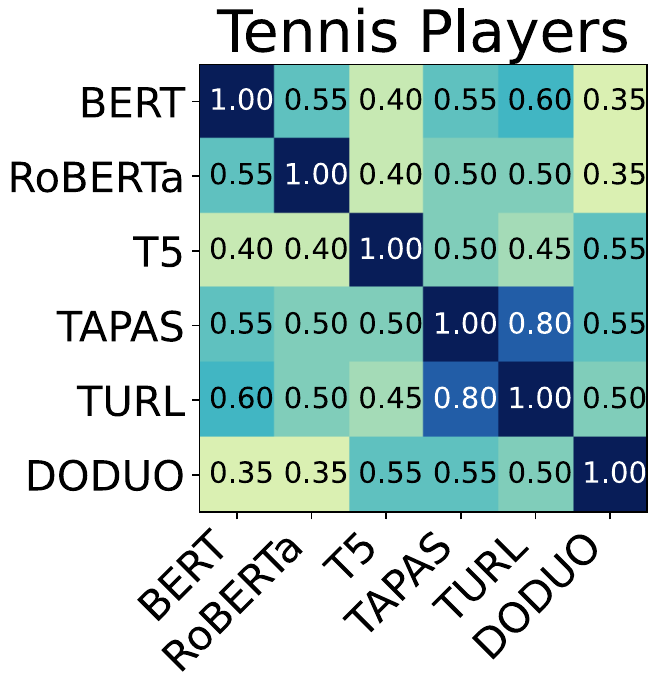}
    \endminipage\hfill
    \minipage{0.305\columnwidth}
      \includegraphics[width=\textwidth]{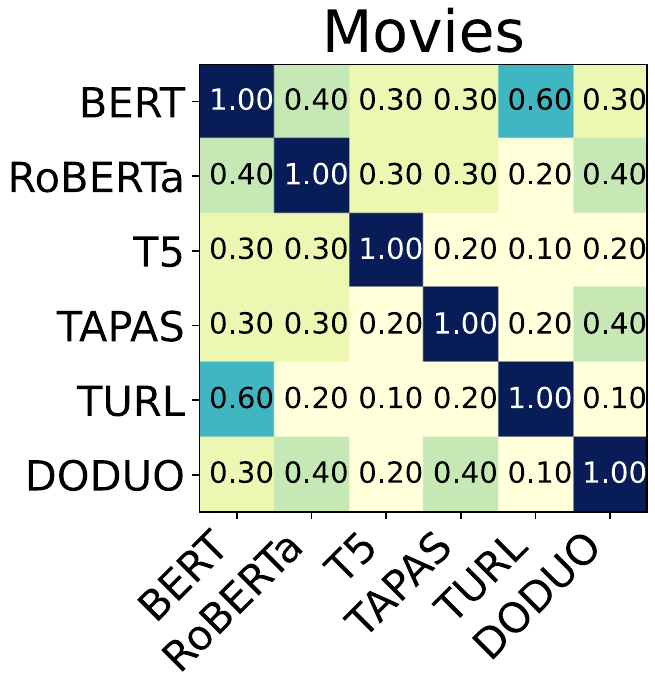}
    \endminipage\hfill
    \minipage{0.39\columnwidth}
      \includegraphics[width=\textwidth]{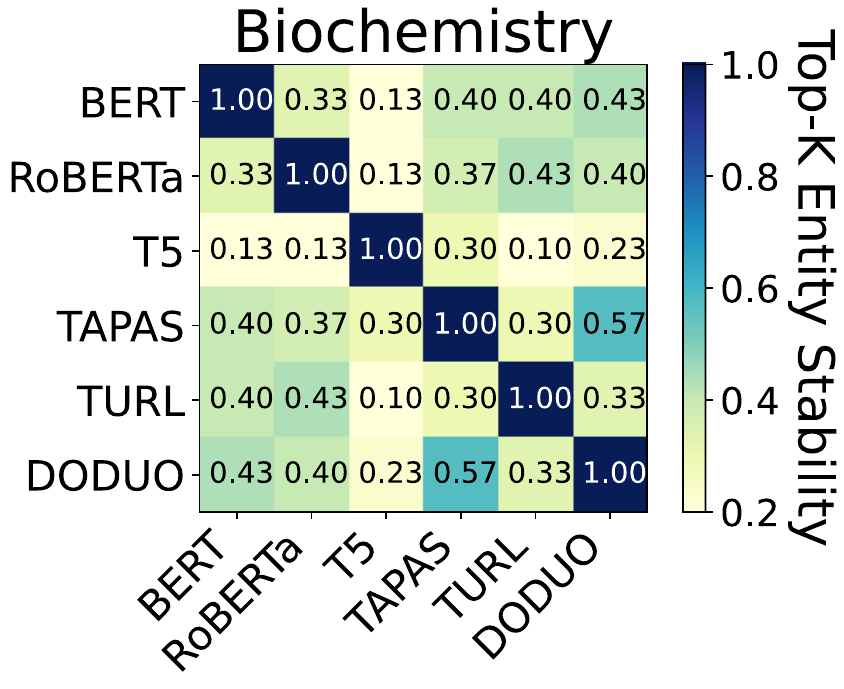}
    \endminipage
    \caption{Pairwise top-10 entity stability with query entities from three distinct domains. Different pairs of models show high entity stability for different domains.}
    \label{fig:entity_stability}
  \end{figure}

  We select query entities from five domains and compare their $K$-nearest neighbors between two embedding spaces: ten greatest men tennis players (Tennis Players), ten most popular movies (Movies), ten most essential nutrients for the body (Biochemistry), ten most valuable technology companies in the U.S., and ten largest countries in the world by area. We plot pairwise average entity stability using heatmaps in Figure~\ref{fig:entity_stability}. Due to space limits, we only show heatmaps of Tennis Players, Movies, and Biochemistry with $K$=10. We observe that domain is a key factor in entity stability. In other words, for different domains, different pairs of models show high entity stability. For instance, \bert and \turl have the highest entity stability for movie entities while \tapas and \doduo have the highest entity stability for biochemistry entities. This suggests for domain-specific tasks, if one finds model A is not feasible, they may want to try model B with relatively lower entity stability with respect to A.
  

\subsection{Perturbation Robustness}\label{sec:res_pr}
  \begin{figure}[!ht]
    \centering
    \minipage{0.49\columnwidth}
      \includegraphics[width=\textwidth]{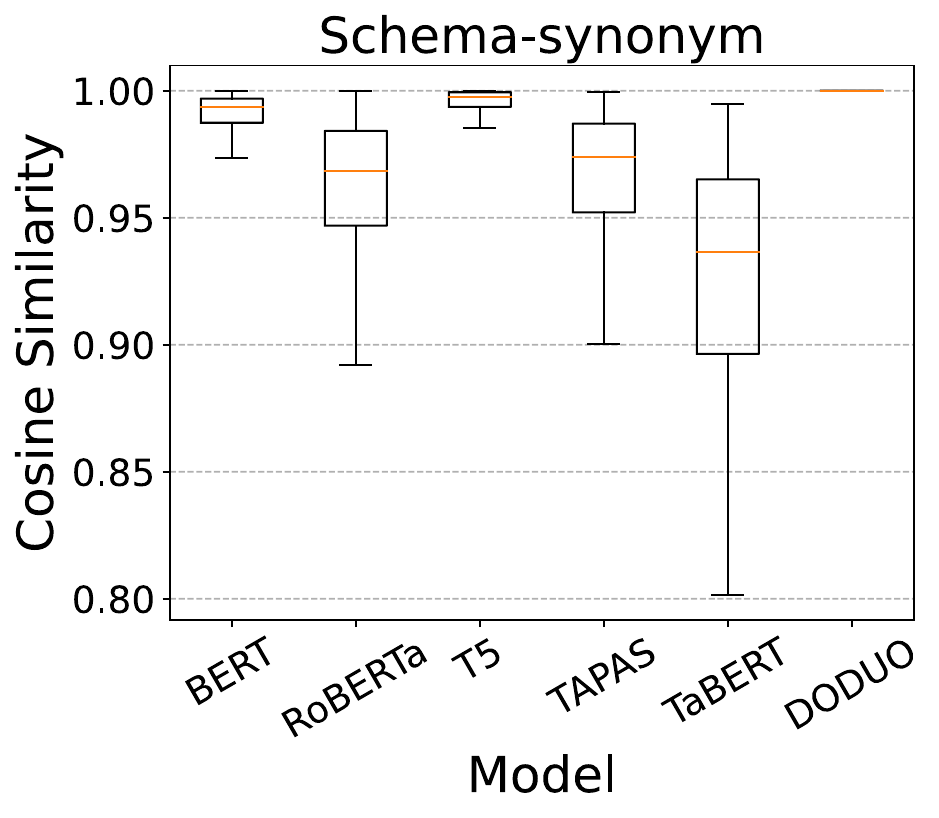}
    \endminipage\hfill
    \minipage{0.49\columnwidth}
      \includegraphics[width=\textwidth]{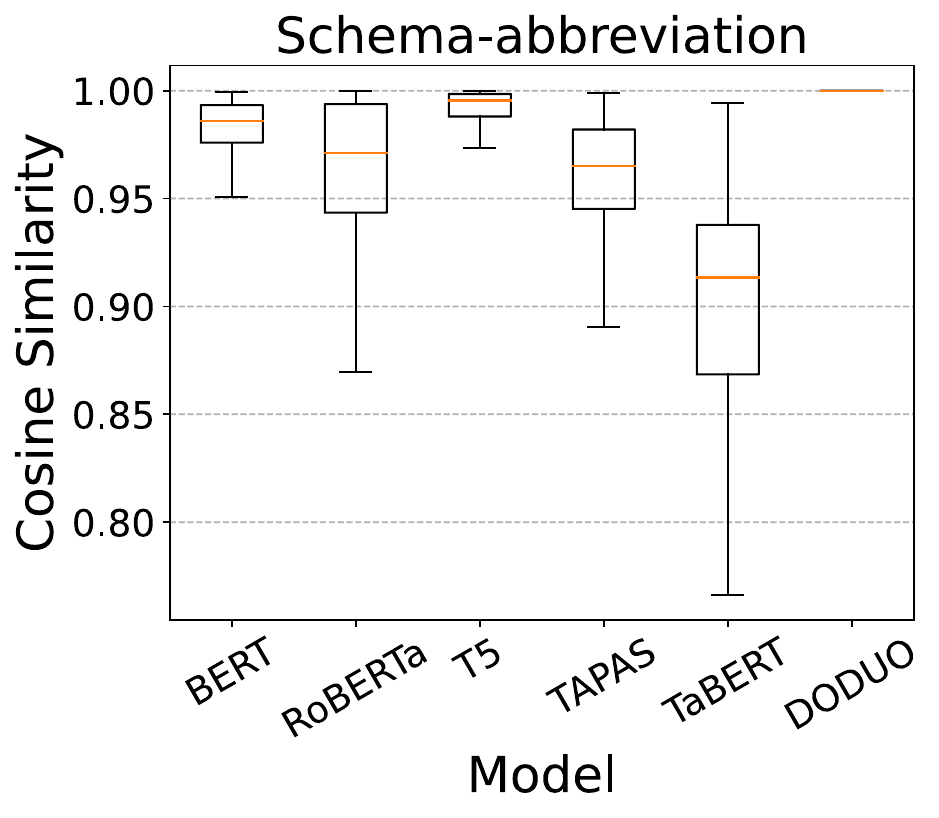}
    \endminipage
    \caption{Distributions of embedding cosine similarities between original columns and perturbed columns. Perturbations only to schemas cause relatively small changes in cosine similarity (except for \tabert).}
    \label{fig:pt_rb}
  \end{figure}

  Figure~\ref{fig:pt_rb} shows distributions of embedding cosine similarities between pairs of an original column and a corresponding perturbed column. Even though both types of perturbations are at the schema level and data values remain unchanged, models exhibit different degrees of robustness, especially in terms of the spread and skewness of the distribution. Vanilla language models \bert and \tfive are most robust to schema-level perturbations with first quartile above 0.97 while entire distributions are above 0.90. Despite being a language model, \roberta surprisingly shows a larger spread with outliers down to 0.75 in synonym perturbations and to 0.65 in abbreviation perturbations. On the table model side, \tabert is least robust to perturbations with the lowest median and first quartile among all models. In contrast, \tapas is more robust with first percentile near 0.95 for both perturbations while it shows relatively large variance as well. \doduo does not show any variance because \doduo only takes in data values for representation inference and simply ignores changes to the schema. Overall, table embedding models in comparison are more sensitive to schema perturbations as they explicitly model the header component of tables and distinguish between headers and data values in representation learning. 

\subsection{Heterogeneous Context}
  \begin{table}[ht!]
    \centering
    \caption{Summary statistics (min, median, and max) of cosine similarities between single column embeddings and contextual embeddings for non-textual and textual data types, on the first and second row, respectively. Incorporating context, especially the entire table, can change column embeddings significantly w.r.t cosine similarity (highlighted in bold).}
    \label{tab:exp_context_heterog}
    \setlength{\tabcolsep}{2pt}
    \setlength{\extrarowheight}{-2pt}
    \resizebox{\columnwidth}{!}{%
    \begin{tabular}{lccc}
    \toprule
    \multicolumn{1}{l}{\textbf{Model}} &
      \textbf{Subject Column} &
      \textbf{Neighboring Columns} &
      \textbf{Entire Table} \\ \midrule
    BERT &
      \begin{tabular}[c]{@{}c@{}}0.72 / 0.89 / 0.99\\ 0.72 / 0.93 / 1.00\end{tabular} &
      \begin{tabular}[c]{@{}c@{}}0.62 / 0.86 / 0.99\\ 0.64 / 0.88 / 1.00\end{tabular} &
      \begin{tabular}[c]{@{}c@{}}\textbf{0.57} / 0.78 / 0.96\\ \textbf{0.51} / 0.79 / 0.99\end{tabular} \\ \midrule
    RoBERTa &
      \begin{tabular}[c]{@{}c@{}}0.76 / 0.83 / 0.89\\ 0.76 / 0.83 / 0.90\end{tabular} &
      \begin{tabular}[c]{@{}c@{}}0.71 / 0.82 / 0.93\\ 0.74 / 0.83 / 0.92\end{tabular} &
      \begin{tabular}[c]{@{}c@{}}0.75 / 0.84 / 0.92\\ 0.76 / 0.85 / 0.93\end{tabular} \\ \midrule
    T5 &
      \begin{tabular}[c]{@{}c@{}}0.77 / 0.85 / 0.93\\ 0.75 / 0.83 / 0.92\end{tabular} &
      \begin{tabular}[c]{@{}c@{}}0.75 / 0.88 / 0.97\\ 0.75 / 0.88 / 0.98\end{tabular} &
      \begin{tabular}[c]{@{}c@{}}0.74 / 0.83 / 0.92\\ 0.75 / 0.83 / 0.98\end{tabular} \\ \midrule
    TAPAS &
      \begin{tabular}[c]{@{}c@{}}0.68 / 0.84 / 0.95\\ 0.52 / 0.83 / 0.98\end{tabular} &
      \begin{tabular}[c]{@{}c@{}}0.58 / 0.80 / 0.97\\ 0.50 / 0.80 / 0.98\end{tabular} &
      \begin{tabular}[c]{@{}c@{}}\textbf{0.35} / 0.64 / 0.92\\ \textbf{0.31} / 0.67 / 0.92\end{tabular} \\ \midrule
    TaBERT &
      \begin{tabular}[c]{@{}c@{}}0.94 / 0.97 / 1.00\\ 0.90 / 0.98 / 1.00\end{tabular} &
      \begin{tabular}[c]{@{}c@{}}0.93 / 0.97 / 1.00\\ 0.89 / 0.97 / 1.00\end{tabular} &
      \begin{tabular}[c]{@{}c@{}}0.89 / 0.95 / 0.99\\ 0.83 / 0.96 / 0.99\end{tabular} \\ \midrule
    DODUO &
      \begin{tabular}[c]{@{}c@{}}0.25 / 0.62 / 0.99\\ 0.34 / 0.80 / 0.99\end{tabular} &
      \begin{tabular}[c]{@{}c@{}}0.14 / 0.59 / 0.99\\ 0.26 / 0.78 / 0.98\end{tabular} &
      \begin{tabular}[c]{@{}c@{}}\textbf{0.06} / 0.45 / 0.87\\ \textbf{0.01} / 0.61 / 0.98\end{tabular} \\ \bottomrule
    \end{tabular}%
    }
  \end{table}

  For both non- and textual data types, we infer from each model column embeddings using only the columns themselves, and adding 1) subject columns; 2) immediate neighbor columns; and 3) the entire tables as context respectively. We compute the cosine similarity between corresponding pairs of single column embeddings and contextual embeddings and show their three-number summary in Table~\ref{tab:exp_context_heterog}. Unsurprisingly, adding different contexts to the inputs changes the embeddings to various degrees. For non-textual columns, among three context settings, models except \doduo preserve high cosine similarity when having subject columns as context (e.g., the median number of \tabert is above 0.96 and that of \bert is close to 0.9) while they (except \tabert) preserve relatively low cosine similarity when having the entire tables as context (e.g., the median number of \tapas is below 0.65). We observe that \tabert embeddings are insensitive to context (the median number is above 0.95 in all three settings) whereas \doduo embeddings are more sensitive to context (the median number is below 0.5 when having the entire tables as context and around 0.6 in the other two settings). We see a consistent trend for textual data. This can have implications that \tabert may not be a good choice for context sensitive downstream tasks and a user may want to try both single column embeddings and contextual embeddings when using \doduo.

\section{Connection to Downstream Tasks}\label{sec:conn}
  From the model characterization through their embedding representations as per the eight properties P\ref{prop:row_order}-\ref{prop:context}, we deduce below the model behaviors on downstream tasks. We illustrate three connections with experimental findings.

  \paragraph{Column Type Prediction (P\ref{prop:row_order}/P\ref{prop:col_order})} In the experiments, \doduo is found sensitive to row/column shuffling and sampling, which are indicators of unstable predictions of \doduo over shuffled data in downstream tasks. To investigate this hypothesis, we randomly sample 1,000 tables from the WikiTables dataset used in the experiments and employ \doduo to predict semantic column types for all columns. For each table, we consider at most 1,000 distinct row-wise permutations for computational efficiency and keep track of how many predictions change per permutation relative to the original order. We find that, over this subset of tables with 5.8 columns on average, $34.0\%$ of the permutated tables yield at least 1 changed column type prediction (averaged over all permutations). $12.8\%$ of the tables have at least 2 changed type predictions while $5.4\%$ of tables have at least 3 changed type predictions.

  \paragraph{Join Discovery (P\ref{prop:sample_fidelity})} \tfive exhibits high sample fidelity even when the sample ratio is low, leading us to anticipate \tfive to be sample efficient in downstream tasks. We implement \tfive in the task of join discovery following the approach and setup in~\cite{DBLP:conf/cidr/CongGFJD23}. Over the NextiaJD testbeds, sampled \tfive embeddings obtain comparable precision and recall as those from full values while the indexing time and lookup time are significantly faster. For instance, on NextiaJD-XS with a sample size of 100 (which is about 5\% of the average number of rows in NextiaJD-XS), there is less than $\pm 3\%$ variation in precision and recall between sampled \tfive embeddings and full-value \tfive embeddings. But the indexing time of using sampled values is more than 7x faster and the lookup time is more than 2$\times$ faster.
  
  \paragraph{Table Question Answering (P\ref{prop:pt_rb})} The task of table question answering (TableQA) refers to answering natural language questions based on information from given tables. In our experiments of the Perturbation Robustness property (Section~\ref{sec:res_pr}), we found that \tapas, among other models, was sensitive to semantics-preserving perturbations to the table schema. Based on this observation, we hypothesize that \tapas may suffer performance degradation on perturbed tables in downstream tasks, such as TableQA, for which it is designed. As anticipated, the TableQA accuracy of \tapas under synonym- and abbreviation perturbation drops by 6.2 and 8.3 points respectively on WikiTableQuestions~\cite{DBLP:conf/acl/PasupatL15}, and 19.0 and 22.2 points respectively on WikiSQL~\cite{DBLP:journals/corr/abs-1709-00103} (see Table 2 and 7 in~\cite{DBLP:conf/acl/ZhaoZNQZTMR23}).
  
  We emphasize that, despite we focus on the characterization of pretrained models (e.g., pretrained version of \tapas), our hypotheses predicated on such characterization propagate to finetuned models (in this case, \tapas models fine-tuned for TableQA).

  \paragraph{Additional Connections} Beyond the three empirically supported anticipations of model behaviors on downstream tasks, we also deduce informed expectations listed below as a result of the characterizations obtained with \sysname for the other properties. This list is not exhaustive as the connection between model characteristics and downstream tasks is not a one-to-one relationship.
  
  \begin{itemize}[leftmargin=15pt]
    \item[\textbf{P\ref{prop:join}}] Low Spearman's coefficient between containment and embedding cosine similarity (e.g., \bert) $\rightarrow$ Join discovery: the containment-based method will complement the embedding-based method in finding join candidates.
    \item[\textbf{P\ref{prop:fd}}] Not preserving functional dependencies $\rightarrow$ Data imputation: imputed values may not maintain functional dependencies between attributes.
    \item[\textbf{P\ref{prop:entity_stb}}] Relative to model A, model B has a lower entity stability than model C $\rightarrow$ Entity retrieval: model B will return fewer entities in common with model A than with model C.
    \item[\textbf{P\ref{prop:context}}] Insensitive to context change (e.g., \roberta) $\rightarrow$ Join discovery: candidates found by single-column and contextual embeddings will largely overlap.
  \end{itemize}

\section{Discussion}\label{sec:discussion}

\paragraph{Impact of Tables with Large Dimensionality} To assess the effect of table dimensions, we examine \bert and \tapas regarding row- and column order insignificance on the NextiaJD-S dataset, averaging over 209k rows and 56 columns. No significant differences emerge on NextiaJD-S compared to tables from WikiTables. The partitioning of large tables into smaller ones, with aggregated embeddings, aligns with our practice for smaller tables.


\paragraph{Limitations} While \sysname encompasses crucial properties for various applications, implementing measures for all properties is unfeasible. For instance, assessing latent topics in tables, vital for retrieval tasks, lacks established measures and appropriate evaluation datasets. The challenge of evaluating models' ability to capture signals across diverse data types, from numeric to textual, persists. Our model analysis is constrained to a representative selection, driven by the availability of code and pretrained model weights. However, \sysname is extensible and open-sourced for analyzing additional models. We acknowledge the potential for future investigation into relationships between property metrics. Our analysis, like any empirical work, is subject to the inherent limitations of dataset specificity. Despite these considerations, we initiate the process of characterizing and understanding embedding representations across relational tables.

\section{Conclusion}

  We introduce \sysname, a downstream-task agnostic analysis framework for table embeddings, gauging the alignment of pretrained embeddings with key relational data model and data distributions properties. Our assessment of nine language- and table embedding models reveals diverse capabilities of different models. Notably, some properties of the relational model and data distributions are not consistently reflected in table embeddings. \sysname provides a valuable tool for guiding model selection in various applications, aiding researchers in model evaluation, and informing future research on novel architectures for tabular data.

\begin{acks}
  This research is supported in part by NSF grants 1946932 and 2312931, by Dutch Research Council (NWO) through grant \\MVI.19.032, and through computational resources and services provided by Advanced Research Computing at the University of Michigan, Ann Arbor.
\end{acks}


\bibliographystyle{ACM-Reference-Format}
\bibliography{sections/reference}

\end{document}